\documentclass[hidelinks,onefignum,onetabnum]{siamart250211}

\usepackage{comment}
\usepackage{lipsum}
\usepackage{amsfonts}
\usepackage{graphicx}
\usepackage{epstopdf}
\usepackage{xspace}
\usepackage{algorithmic}
\usepackage{todonotes}

\ifpdf
  \DeclareGraphicsExtensions{.eps,.pdf,.png,.jpg}
\else
  \DeclareGraphicsExtensions{.eps}
\fi

\newsiamremark{example}{Example}
\newsiamremark{observation}{Observation}
\newsiamremark{remark}{Remark}
\newsiamremark{hypothesis}{Hypothesis}
\newsiamthm{claim}{Claim}
\newsiamthm{property}{Property}

\headers{On Computing the Dollo-1 phylogeny in polynomial time}{P. Bonizzoni, G. Della Vedova, M. Soto Gomez, and G. Trucco}

\title{Computing the Dollo-1 phylogeny  in polynomial time\thanks{Submitted to the editors DATE.
\funding{This project has received funding from the European Union’s Horizon 2020 Research and Innovation Staff Exchange programme under the Marie Skłodowska-Curie grant agreement No. 872539}}}

\author{Paola Bonizzoni\thanks{Dipartimento di Informatica Sistemistica e Comunicazione, 
Università degli Studi di Milano--Bicocca  
  (\email{paola.bonizzoni@unimib.it}).}
\and Gianluca Della Vedova\thanks{Dipartimento di Informatica Sistemistica e Comunicazione, Università degli Studi di Milano--Bicocca  
  (\email{gianluca.dellavedova@unimib.it}).}
\and Mauricio Soto Gomez\thanks{Dipartimento di Informatica, Università degli Studi di Milano  
  (\email{mauricio.soto@unimi.it}).}
\and Gabriella Trucco\thanks{Dipartimento di Informatica, Università degli Studi di Milano 
  (\email{gabriella.trucco@unimi.it}).}
  }

\usepackage{amsopn}

\newcommand{\hgraph}{{hybrid-graph}\xspace}
\newcommand{\scanonical}{{3-canonical}\xspace}

\newcommand{\degenerate}{{degenerate}\xspace}
\newcommand{\adegenerate}{{almost degenerate}\xspace}
\newcommand{\sactive}{{pair-active}\xspace}

\newcommand{\simple}{{complete}\xspace}
\newcommand{\incomplete}{{incomplete}\xspace}
\newcommand{\grb}{\ensuremath{G_{RB}}\xspace}

\newcommand{\I}{\ensuremath{\mathcal{I}}}
\newcommand{\A}{\ensuremath{\mathcal{A}}}
\newcommand{\IG}{\ensuremath{\mathcal{I}(G_{RB})}\xspace}

\newcommand{\AG}{\ensuremath{\mathcal{A}(G_{RB})}\xspace}

\newcommand{\grbM}{\ensuremath{G_{RB}^{M}}\xspace}
\newcommand{\grbm}{\grbM}
\newcommand{\grbe}{\ensuremath{G_{RB}^{A}}\xspace}
\newcommand{\grbp}{\ensuremath{G_{RB}^*}\xspace}

\newcommand{\rs}{red $\Sigma$-graph\xspace}

\newcommand{\Hd}{\ensuremath{\cal P}\xspace}

\newcommand{\free}{red universal\xspace}

\newcommand{\ie}{\emph{i.e.},\xspace}

\newcommand{\dolloone}{Dollo-$1$\xspace}

\newcommand{\pendant}{minimal \xspace}
\usepackage{subcaption}
\usepackage{forest}
\usepackage{xspace}

\newcommand{\notaestesa}[2]{%
  {\color{red!75!black}%
    [\,\textbullet\,\textsf{\textbf{#1:}} %
    \textsf{\footnotesize#2}\,\textbullet\,]}%
}

\newcommand{\pb}[1]{\notaestesa{PB}{#1}}

\ifpdf
	\hypersetup{
		pdftitle={On Computing the persistent perfect phylogeny in polynomial time},
		pdfauthor={P. Bonizzoni, G. Della Vedova, M. Soto Gomez, and G. Trucco}
	}
\fi

\begin{document}

\maketitle

\begin{abstract}
The Dollo model for reconstructing evolutionary trees from binary characters has been proposed as a generalization of the infinite sites model, also known as the Perfect Phylogeny. In particular, the Dollo model is considered more realistic than the Perfect Phylogeny for inferring the evolution of tumor mutations.

In the case of binary matrices, the Dollo-$k$ model requires an evolutionary tree in which each character ---corresponding to a column in the input matrix --- may change from $0$ to $1$ at most once, and from $1$ to $0$ at most $k$ times throughout the entire tree.

Given a binary matrix, the problem of deciding whether there exists a Dollo-$k$ tree compatible with the matrix is NP-complete for any fixed $k \geq 2$, while computing a Dollo-$0$ tree corresponds to the Perfect Phylogeny decision problem, which admits a simple linear-time algorithm.

The Dollo-$1$ tree problem corresponds to the Persistent Phylogeny problem, whose computational complexity --- albeit under an equivalent formulation --- was posed as an open question 20 years ago.

We solve this problem by presenting a polynomial-time algorithm for the Persistent Phylogeny problem. Our solution relies on efficiently solving a specific class of binary matrices, represented as bipartite graphs called \emph{skeleton graphs}, or simply skeletons. In these graphs, characters are  \emph{maximal}, that is their corresponding sets of species are not related by inclusion.
\end{abstract}



\section{Introduction}
The problem of reconstructing an evolutionary history from characters  is a classical topic in computational biology~\cite{Gusfield,semple2003phylogenetics}.

The first and most widely studied formulation of the problem is the \emph{Perfect Phylogeny} problem. In this model—as in this paper—the input is a binary ($0$/$1$) matrix, where the rows represent a set of species (or taxa), the columns represent a set of characters, and the matrix entries indicate which characters are present in each species. Specifically, a value of $0$ denotes the absence of a character, and $1$ denotes its presence in a given taxon.

The goal is to construct a phylogeny—that is, a tree in which the leaves (and possibly some internal nodes) are labeled by the species (rows)—such that each edge corresponds to a mutation event: a change in the state of a character. A change from $0$ to $1$ represents a character gain, while a change from $1$ to $0$ represents a character loss, modeling the evolution of ancestral taxa.

In the undirected Perfect Phylogeny problem, the root of the tree may initially possess characters in state $1$, which can each transition to state $0$ (i.e., be lost) exactly once in the tree. Similarly, characters that begin in state $0$ at the root can change to state $1$ exactly once in the tree. When the input matrix is complete, the problem can be solved in linear time using a well-known algorithm~\cite{Gus91,Gusfield}.

In the directed version of the problem, all characters are assumed to be in state $0$ at the root—thus, the Directed Perfect Phylogeny problem forbids character losses entirely. A linear-time algorithm for this version, even on incomplete matrices, was presented in~\cite{DBLP:journals/corr/abs-2010-05644}, improving upon earlier work in~\cite{Sha}.

The Perfect Phylogeny model has found numerous applications, primarily in bioinformatics—particularly in haplotyping~\cite{Gus2,Gus06,Boniz,Gus02}, protein domain analysis~\cite{przytycka2006graph}, and, more recently, tumor phylogeny inference~\cite{hajirasouliha14_recon_mutat_histor_multip_sampl,hajirasouliha_combinatorial_2014,malikicPhISCSCombinatorialApproach2019,10.1093/bioinformatics/btv261}.

An important generalization of the problem
includes some well-studied relaxation of the perfect phylogeny: the Camin-Sokal and the Dollo model~\cite{caminsokal,Dollo}.
The Camin–Sokal model allows each character to be gained multiple times in different branches of the phylogenetic tree, but no character losses are permitted. Symmetrically, the Dollo model allows each character to be gained at most once, while permitting multiple losses along different lineages.

Recent findings suggest that the Dollo model is well-suited for reconstructing the evolution of genes in eukaryotic organisms, where state changes from $0$ to $1$ and from $1$ to $0$ represent gene gains and losses, respectively~\cite{bonizzoni_when_2014}. This is biologically motivated by the observation that while gene losses are relatively common, independent gains of the same gene in different lineages are rare events~\cite{Dollo}.

Moreover, the Dollo model has recently attracted considerable attention in the context of reconstructing the clonal evolution of tumors.
Recent literature on tumor evolution from single-cell and bulk data extends the classical infinite sites assumption by allowing losses~\cite{bonizzoni2018does,ciccolellaInferringCancerProgression2020,el-kebirSPhyRTumorPhylogeny2018,gpps,vedova17} through the Dollo-$k$ model, where $k$ denotes the total number of losses for each gained character in the tree.

From the computational complexity point of view, determining if a binary matrix has a Dollo-$k$ phylogeny is NP-complete for any fixed $k \ge 2$~\cite{goldberg_minimizing_1996}.
While the Dollo-$0$ corresponds to the directed perfect phylogeny, for which a linear-time algorithm exists~\cite{Gus91,Gusfield}, the Dollo-$1$ problem is still open \cite{goldberg_minimizing_1996,benham1995hen}.
The newly-found relevance of the problem, has led to a deeper investigation.
A phylogeny compatible with the Dollo-$1$ model is also called a \emph{persistent phylogeny}, and a character that has been lost somewhere in the tree is called a persistent character~\cite{przytycka2006graph,DBLP:journals-tcs-BonizzoniBDT12,bonizzoni_when_2014}.
Some practical approaches to the Dollo-$1$ problem have been proposed in the literature, most notably a fixed-parameter algorithm, where the parameter is  the number of characters~\cite{DBLP:journals-tcs-BonizzoniBDT12},  an integer linear programming approach~\cite{gusfield_persistent_2015}, and some variants where mutation losses are constrained~\cite{bonizzoni_explaining_2014,Bonizzoni2016}.
More recently, a different variant of the Dollo-$1$ and the Dollo-$k$ problems, where the topology of the tree is known, has been studied~\cite{WICKE2020102046,bouckaert2020combinatorial}, leading to efficient algorithms and a deeper understanding of the combinatorial properties of those models.
In this paper we provide a polynomial time algorithm to solve  a more general version of the Dollo-$1$ problem that we call \emph{Persistent  Phylogeny  Problem}  as in~\cite{DBLP:journals-tcs-BonizzoniBDT12}, in short PP problem, where characters can be already gained in the root of the tree.
The PP problem is over instances that consist of a graph representation of binary matrices and asks for the construction of a tree, if it exists, that is a persistent phylogeny.

The main contribution of our paper stems from the investigation of the PP problem when restricted to matrices with only \emph{maximal} characters,  a character $c$ is maximal if the set of species having $c$ is not included in the set of species of another character $c'$. This restriction is particularly interesting because the topmost edge of a general persistent phylogeny is always labeled by a maximal character. Furthermore, we represent instances of the PP problem using a special class of bipartite graphs  called \emph{red-black graphs},   , and we show that solving the PP problem is equivalent to recognizing \emph{reducible} red-black graphs—those that can be reduced to an empty graph through a series of graph operations that remove edges and nodes.

\subsection{Structure of the paper}

The paper is organized in PART I: solving skeleton graphs and PART II: solving the general graphs.

We first introduce the notion of \emph{red-black graph}, which serves as our representation of matrices corresponding to instances of the Persistent Phylogeny (PP) problem. In this context, we define the concepts of \emph{realization of characters and realization of species} —two operations on the graph that, when applied iteratively, reduce the graph to the empty graph if and only if the instance admits a persistent phylogeny tree representation. If such a reduction to the empty graph exists, we say that the red-black graph is \emph{solvable}.

We show that there is a bijection between these graph operations and the construction of a tree that solves the instance. Next, we introduce the construction of the Hasse diagram associated with a red-black graph, which is based on the inclusion relationships among the species in the graph. Using this Hasse diagram, we propose an algorithm to solve a special class of graphs containing only maximal characters, which we call \emph{skeleton graphs}. Recall that a character is defined as maximal if its set of species is not contained within the set of species of any other character.

Indeed, in Part I, we present an algorithm for solving skeleton graphs based on iteratively realizing a \emph{safe species} —that is, a species that can be shown to correspond to a source of a chain in the Hasse diagram associated with the graph.

The characterization of solutions for skeleton graphs concludes Part I. The trees that solve these graphs exhibit a very special structure.

Part II is centered on applying an \emph{invariant property} for general graphs: any reduction of such graphs always begins with a maximal character of the corresponding skeleton graph. Leveraging this invariant property, we propose a polynomial-time algorithm for solving general graphs.

\section{Preliminaries}
\label{sec:preliminaries}

The classical representation of an instance of the phylogeny reconstruction problem is a $n\times m$ binary matrix $M$ where rows and columns of $M$ are associated with a set $S$ of species and a set $C$ of characters respectively; then species $s\in S$ has the character $c\in C$ if and only if the entry $M[s,c]$ is equal to one.

Observe that an alternative representation of an instance of the phylogeny reconstruction problem   is given by    bipartite graphs consisting of two  vertex sets $(C,S)$, associated with the species set $S$ and character set $C$ respectively, while an edge between a character node and a species node represents the presence of the character in the species. Figure~\ref{fig:instance-extended} depicts an example of a binary matrix and its corresponding bipartite graph representation.

\subsection{Red-black graphs}
\label{sec:rbg}

\begin{figure}[tb]
	\centering
	\includegraphics[width=1.0\linewidth]{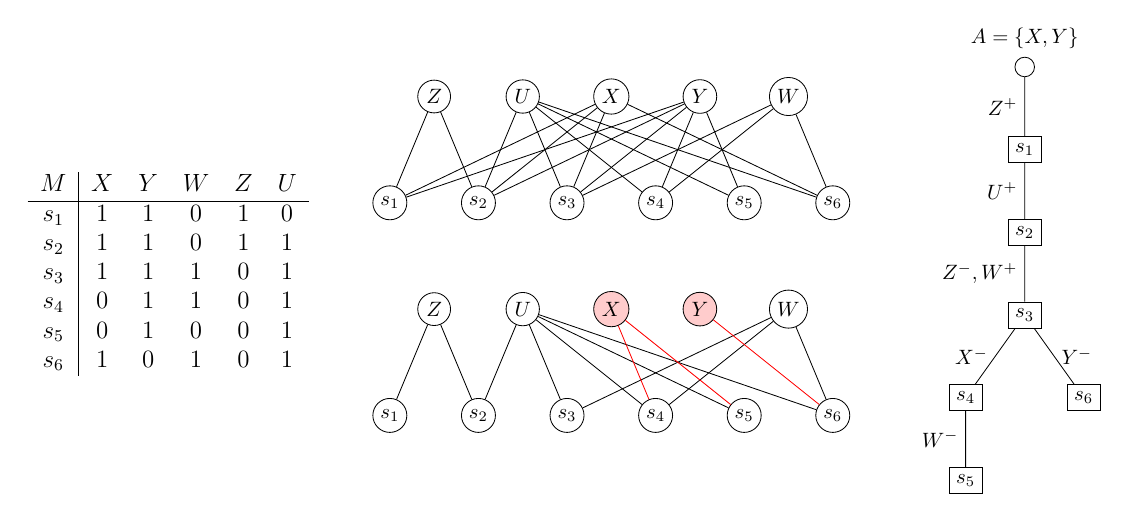}
	\caption{A  graph representation (top-center) of characters and species given by the red-black graph. The red-black graph associated with the set of active characters $A=\{X,Y\}$ (bottom-center), and a tree solving  the graph (right). 
	}
	\label{fig:instance-extended}
\end{figure}

In order to deal with phylogenetic trees where characters can be lost, in this work we consider a special graph representation of binary matrices introduced in~\cite{DBLP:journals-tcs-BonizzoniBDT12}, called \emph{red-black} graph and denoted as \grb.
This general representation allows us to deal with a  set of characters that were already gained in a root or a node of the phylogenetic tree, and that can be later lost, if they are persistent in the tree. Indeed,  observe that the main idea in~\cite{DBLP:journals-tcs-BonizzoniBDT12} is that each node of the phylogenetic tree is associated with   a red-black graph: this graph represents the subtree rooted in the node (Figure~\ref{fig:alberoesempio}).
Similarly to the case of the phylogeny reconstruction problem with a  root where all characters have state zero,   the red-black graph \grb is a bipartite graph having two sets $C$ and $S$, where $C$ denotes the set of characters and $S$ denotes the set of species in the input.
Edges, on the other hand, are either black or red, to distinguish between the two possible states of a character in the current node of the tree  (see Figure~\ref{fig:instance-extended}  and  Figure~\ref{fig:alberoesempio}): \emph{active} (i.e.  gained in the node) and \emph{inactive} (i.e. not yet gained in the node).
More precisely, the main idea of a graph rapresentation associated to a binary matrix $(M,A)$ consisting of species and  characters over binary states, is that a black edge connects character $c$ to species $s$ whenever the state of  $c$ is $1$ in species $s$ and $c$ is inactive, equivalently $M[c,s]=1$ and $c$ is not in $A$, meaning that it will be gained in the tree. If instead the character $c$ is already active in species $s$, equivalently $M[c,s]=1$ and $c\in A$, then $c$ is not connected to $s$, and moreover $c$ is connected by red-edges to those species $s'$ where $c$ has state $0$ (i.e. $M[c,s']=0$), meaning that $c$ will be lost in such species, i.e. it is persistent in such species.
Then, in a connected graph \grb, an  inactive character $c$ is adjacent via black edges to all species $s$   that  have character $c$,  while an active character $c$ is adjacent via red edges to species $s$ that do not have character $c$.  Moreover, an inactive character may become active, as we will show later in the paper (see also Figure~\ref{fig:instance-extended} for an example).

More formally, we give the following  definition of red-black graph.

\begin{definition}[Red-black graph for $A$]
	\label{def:red-black}
    The \emph{red-black} graph with a set $A$ of active characters is a bipartite graph   $\grb=(S\cup C, E(\grb))$ with two sets of nodes, a set $S$ of species and a set $C$ of characters,  and  edges $E$ connecting a species in $S$ to a character $c$ in $C$, where we distinguish  $E_B$ the set of \emph{black edges} and   $E_R$ the set of \emph{red edges}, that is $ E(\grb) = E_R \cup E_B$ such that each character $c \in C$, with $c \in A$ is incident only to red edges or to black edges,   if $c \not\in A$.
\end{definition}

Given a vertex $v$ of \grb, then $N(v)$
is the set of neighbors of the vertex $v$ in \grb, i.e. the set of vertices to which $v$ is connected.
We denote by $S(c)$
the set of species that have the character $c$, that is
$S(c)=N(c)$ if $c$ is inactive and $S(c)=V\setminus N(c)$ if $c$ is active.
Informally, the set of species that have character $c$ in graph \grb are  those species for which the state vector  has character $c$ at $1$.
Similarly, we define  the set $C(s)$ of characters that the species $s$ has:
$C(s)=\{c \in C: s\in S(c)\}$.
Moreover, given two characters $c_{1}$ and $c_{2}$, we will say that $c_{1}$ \emph{includes} $c_{2}$ if $S(c_{1})\supseteq S(c_{2})$.

In fact, an active character is assumed to be gained in the root of the tree
represented by the graph,  as we will detail in the definition of persistent
phylogeny.

\begin{definition}[\pendant species]
    \label{def:pendant}
    Given a species $s$ of \grb, then $s$ is \emph{\pendant} if there is no other species $s'$ such that $s'$ is incident to a subset of characters of $s$.
\end{definition}


By the above definition  \ref{def:red-black}, the input of the problem of reconstructing a phylogenetic tree can be also given in the form of a red-black graph.

\begin{example}
	An example of active characters is represented by the red characters $X$  and $Y$ in the bottom center graph of Figure~\ref{fig:instance-extended}: indeed $X$ is  connected by red egdes to species $s_4$ and $s_5$ that are not in $S(X)$ and similarly $Y$ is connected by red edges to species $s_6$ which is the only one not in $S(Y)$.
	Instead $X$ and $Y$ are inactive in the upper center graph of Figure~\ref{fig:instance-extended}.
\end{example}

Given a red-black graph \grb, $\AG$  and $\IG$ are respectively the set of
\emph{active} (respectively \emph{inactive})   characters of \grb;
when \grb is clear from the context, we will omit it, writing simply $\A$ and $\I$.

\begin{definition}[Universal character]
A character $c$ is \emph{\free} (respectively
universal) in $\grb$, if $c$ is adjacent via red (respectively black) edges to
all species belonging to the same connected component of character $c$.
\end{definition}

\subsection{The Persistent Phylogeny problem}

In its standard version, the Dollo-1 problem assumes that the root has no active characters.
We deal with a general version of the problem, where some characters are already
active in the root --- those characters might only be lost.
This notion is formalized below as Definition~\ref{def:persistent-perfect-phylogeny}. Observe that the Definition~\ref{def:persistent-perfect-phylogeny} is given for an instance consisting of a red-black graph, but it can be equivalently given for a binary matrix of characters and species.

\begin{figure}[htb!]
	\centering
	\includegraphics[width=1.0\linewidth]{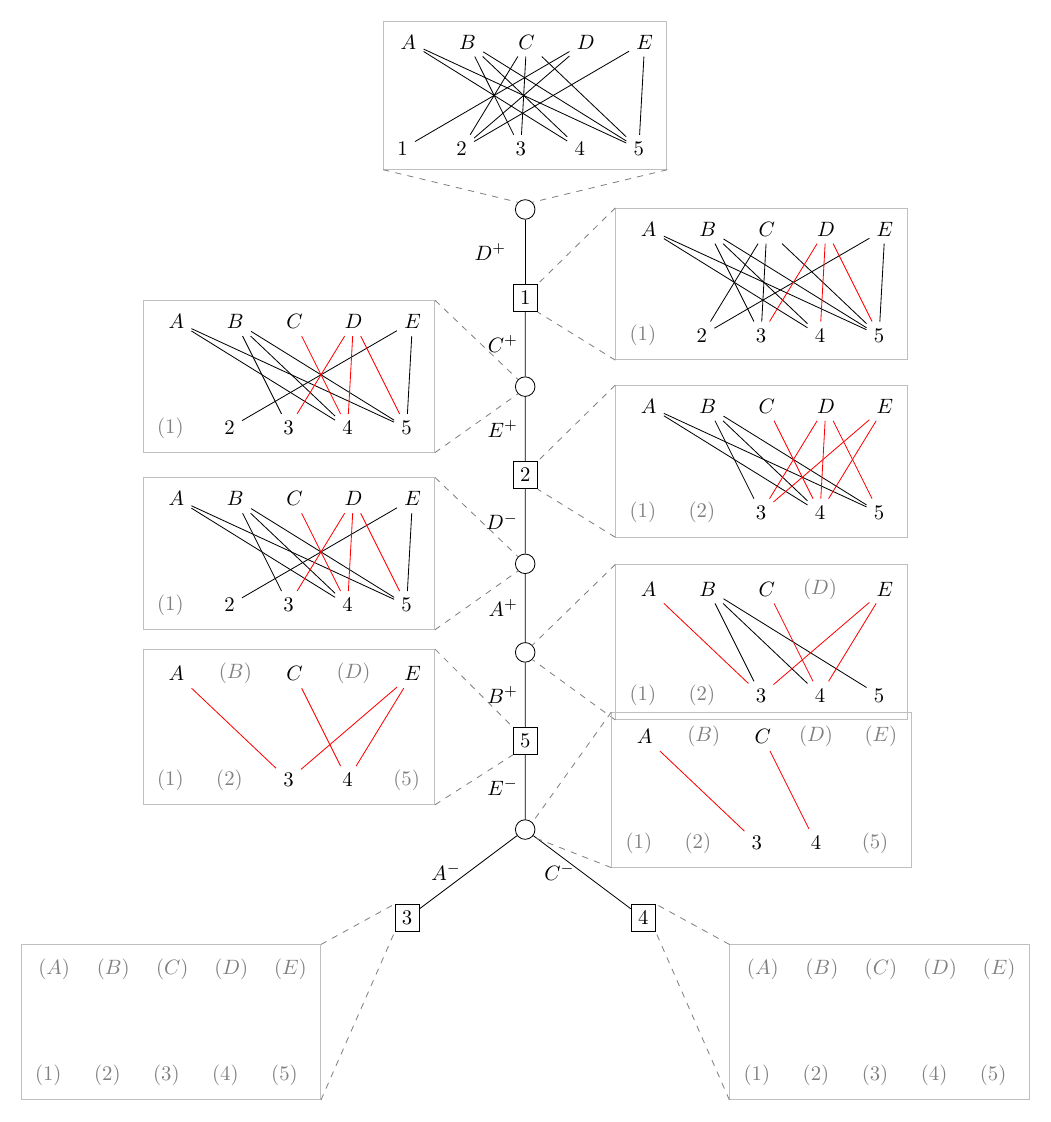}
	\caption{Example of a \dolloone instance and its solution.
		The instance is the red-black graph labeling the root of the tree, while the
		tree is the solution of the instance.
		Each node of the tree is labeled by the red-black graph resulting from the realization of
		a single character on the graph labeling its parent, where numbers represent
		species and letters represent characters.
		Gray vertices that are between parenthesis are not actually part of the graph.
		Moreover, the subtree rooted at a given node is a solution of the red-black
		graph labeling such node.
		The tree is built from the realization of the sequence $\langle D,C,E,A,B\rangle$ of
		inactive characters.
		The tree is not in its standard form. Notice that isolated vertices represent
		the realization of the species and their  characters and they will be removed from the graph.
	}
	\label{fig:alberoesempio}
\end{figure}

In order to define the Persistent Phylogeny tree solving a graph, we need to define the notion of \emph{state vector} of a species $s$ in  $\grb$.  
Given   $\grb$   a red-black graph over a set $C$ of $m$ characters and a set $S$ of $n$ species, then the \emph{state vector} of a species $s$ in \grb is the $m$-vector $l_s$ where $l_s[j]=1$ if  character $c_j$  is inactive and species $s$ has a black edge incident from character $c_j$  or  $c_j$ is active and species $s$  has no incident red edge from $c_j$. In all other cases $l_s[j]=0$.
For example, observe that in Figure~\ref{fig:alberoesempio}, the state vector for species $s_5$ has $1$ only for the inactive character $U$ and the active character $Y$.
By the above definition, the state vector of a species $s$ in the graph corresponds to the row $s$ of the binary matrix associated to the red-black graph.

\begin{definition}[Persistent Phylogeny or general Dollo-1]
	\label{def:persistent-perfect-phylogeny}
	Let $\grb$ be a red-black graph over a set $C$ of $m$ characters and a set $S$ of $n$ species. Given a set $\A$ of active characters, with $\A \subseteq C$, the \emph{Persistent Phylogeny} problem, in short \emph{PP} problem, for the graph  $\grb$ asks for a rooted tree $T$, called \emph{persistent phylogeny} such that:
	\begin{enumerate}
		\item each node $x$ of $T$ is labeled by a binary vector $l_x $  of length $m$,  such that $l_x[j] =1$ or $l_x[j] =0$, if character $c_j$ has state $1$ or $0$, respectively in node $x$;

		\item
		      the root $r$ of $T$ is labeled by a vector $l_r$ such that $l_r[j]=1$ if and only if $c_j \in A$;
		\item
		      each edge $e=(x,y)$ is labeled by a positive signed character,  i.e. $c_j^{+}$, meaning that  $c_j$ is gained, if  $l_x[j]=0$ and $l_y[j]=1$,
		      otherwise if $l_x[j]=1$ and $l_y[j]=0$ the label of the edge is a negative character i.e.
		      $c_j^{-}$, meaning that  $c_j$ is lost;
		\item
		      each character is gained and lost at most once in the tree;

		\item
		      for each species $s$ of $\grb$ there exists a node $x$ of $T$ such that the vector $l_{x}$ is equal to the state vector for $s$. In this  case, $s$ labels the node $x$.
	\end{enumerate}
	Then we say   that graph \grb is solved by the tree $T$.
\end{definition}

Intuitively, a tree $T$ solving a graph \grb is just a tree where the state vector of the root has $1$ value only for active characters and each character is gained or lost at most once in the tree $T$ in such a way that the state vector of all species of the graph are associated to nodes of the tree.
Since the graph is an alternative representation of  a matrix $(M,A)$, where $A$ is a set of active characters of graph \grb,  the above definition \ref{def:persistent-perfect-phylogeny} can be equivalently given for an input matrix $(M,A)$, where $A$ is a set of active characters. In this last case, for each row $r$ of the matrix, the tree $T$ will have  a species having state vector that  corresponds to the row $r$ of the matrix, while the root of tree $T$ is the state vector with only $1$'s  for the active characters.  

We require all solutions to be standard trees, that is each node, with the
possible exception of the root, is labeled by a species or has at least two
children, as formalized with the following definition.

\begin{definition}[standard form]
	\label{def:standard-form}
	Let $T$ be a tree  solving a red-black graph \grb.
	Then $T$ is {\em standard} if each  node $x$ of the
	tree is labeled by a species of the  red-black graph \grb  or  $x$ has at least two
	children, or $x$ is the root of $T$.
\end{definition}

Given a solution $T$ that is not standard, we can obtain an equivalent standard
solution $T_{1}$ by merging into a single edge all its paths $(x_1, \ldots , x_{k})$ such
that, for $1\le i<k$, $x_{i}$ is not labeled by a species and has exactly one child.
The label of the new edge is the union of the labels of the edges in the
original path.

Observe that in virtue of property $3$ of the previous Definition \ref{def:persistent-perfect-phylogeny}, for each character $c\in C$, 
there exists at most an edge $e=(x,y)$ labeled by a signed character $c_j^{-}$ and $c_j^{+}$.
In the later case, if there exists an edge  $e'=(u,v)$ labeled by $c_j^{-}$ then either $c_j \in A$ and label $c_j^{+}$ does not occur in the tree, otherwise the edge $e$ labeled $c_j^{+}$  and $e'$ occurs along the same path from the root to a leaf of $T$ (clearly $e$ is closer to the root than $e'$ along this path).

We are now able to formalize  the PP problem.

\begin{definition}[Persistent Phylogeny Problem]
\label{def:PP-problem}
Let \grb be a red-black that is associated to a binary matrix $(M,A)$, for $A$ a set of active characters. Then the \emph{Persistent Phylogeny Problem}, in short PP problem, asks for the existance of a persistent phylogeny tree or Dollo-1 for the  graph \grb.
\end{definition}

The main idea behind the red-black graph, as illustrated in~\cite{DBLP:journals-tcs-BonizzoniBDT12}, is that there exists a bijection between graph operations involving character nodes of the graph, called \emph{realization} of characters,  and the construction of the phylogenetic tree in terms of adding characters to the tree and losing characters in the tree, as well as adding species node to the tree.
Figure~\ref{fig:alberoesempio} illustrates graphs associated to nodes of a phylogenetic tree and obtained by updating the initial graph in the root during the addition of characters in the tree.

Observe that while the graph in the root represents the final tree, the graphs in each node of the tree represents  the subtree rooted in that node. Indeed,    the construction of the tree can be expressed by a recursive procedure: this fact has been detailed in~\cite{DBLP:journals-tcs-BonizzoniBDT12} and we use this fact in the paper.

More precisely,  if $T$ is a solution for graph $\grb$ and $T_x$ is the subtree of $T$ rooted at a node $x$, then $T_x$ is a solution for the red-black graph $\grb(x)$ which is obtained from  the graph  $\grb$ by graph operations.   Now,  the relationship between  the graph $\grb$  and graph $\grb(x)$ has already being characterized in ~\cite{DBLP:journals-tcs-BonizzoniBDT12} by using the realization of characters and species (see section~\ref{subs:realization}). 
This property is at the base of our algorithm, as we will detail below.



\begin{remark}
For the sake of simplicity, and without loss of generality, we assume that the instance graph of the problem will not have identical characters or species.
We also assume that red-black graphs are connected and that no node, other than possibly the root, has a label consisting of all zeros.
\end{remark}


\subsection{Realization of characters and species in the red-black graph}
\label{subs:realization}

Let us now describe the graph operations over a red-black graph that are used to characterize the  construction of a tree solving an instance graph~\cite{DBLP:journals-tcs-BonizzoniBDT12}: these operations are called \emph{realizations} and aim  to remove all edges of the input red-black graph.
In the following we show that the \emph{realization of an inactive character $c$} in the red-black graph corresponds to adding the positive signed character $c$ to the tree  (i.e. $c^{+}$), during its construction. Another operation on the red-black graph is then removing an active   \free character $c$: this operation corresponds to adding to the tree the negative signed character $c$ (i.e. $c^{-}$).

\begin{definition}[realization of an inactive character $c$ or universal character]~\cite{DBLP:journals-tcs-BonizzoniBDT12}
	\label{def:realizing-in-red-black-graph}
	Let $\grb$ be a connected red-black graph and $c$ be an inactive character of $\grb$.
	Let $D(c)$ be the set of species in the connected component of $\grb$ that contains $c$.
	The output of the \emph{realization of $c$} is the red-black graph \grbp obtained from $\grb$ after performing the following sequence of operations:
	\begin{enumerate}
		\item add a red edge between $c$ and each species in $D(c) \setminus N(c)$;
		\item delete all black edges incident on $c$;
		      \item\label{item:free} remove all \free (universal) characters and the edges incident on it, until no \free (universal) character exists in the resulting graph;
		\item remove all  isolated vertices.
	\end{enumerate}
	If the vertex labeled by $c$ has not been removed, the character $c$ becomes an active character in \grbp.
\end{definition}

\begin{figure}
	\centering
	\includegraphics[width=1.0\linewidth]{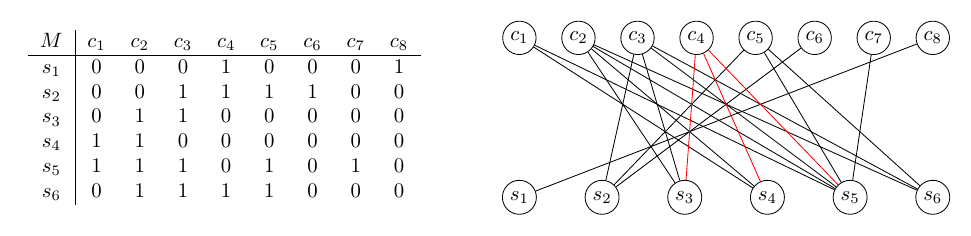}
	\caption{A binary input matrix $(M, \A)$ (left) and its graph representation (top-center) where set $A =\{c_4\}.$}
	\label{fig:red-black-1}
\end{figure}

\begin{example}
    To better illustrate the meaning of realization of a character in the red-black graph instance, consider the matrix $M$ of
    Figure~\ref{fig:red-black-1} representing the incident matrix of a red-black graph with all black edges.  Now, the graph reported in the figure is the one where $c_4$ is realized. Observe that it represents the instance $(M,A)$ where the set $A$ of active characters consists only of character $c_4$.
    Thus red-black graphs instances replace binary matrices with active characters.
\end{example}

Observe that the realization is only defined for inactive characters, therefore red edges can only be removed when they are incident  to a  character that becomes \free as described in step~\ref{item:free} of Definition \ref{def:realizing-in-red-black-graph} and,  by construction, it can be applied only once.
Observe that by definition of realization, isolated vertices will then be removed
from the graph.



\begin{definition}[realization of a species~\cite{Bonizzoni2016}]
	\label{def:species-realization}
    Let $s$ be a species of a red-black graph \grb such that is minimal.
    Then the {\em  realization of  the species  $s$}  on \grb consists of the
    realization  of all inactive characters of \grb that $s$ has, \ie of characters in $C(s)\cap \IG$.
    The realization of a species is
    only defined if $s$ becomes an isolated vertex in the graph, regardless of the order in which characters are realized.
\end{definition}

Observe that the following notion is only defined for \pendant species, as otherwise the ordering of realization of characters may change the  resulting graph.
Indeed, if $s$ is not minimal it means that there exists a subset of characters of $s$ that form a species $s'$ that can be removed from the graph before $s$. Changing the order of characters realized in $s$ may not lead to the removal of $s'$.

Following the same argument, we can generalize the notion of realization to a
sequence of inactive characters, which is the result of realizing in order the
characters of the sequence.
If the result of a  realization is a red-black graph \grb without inactive
characters, the realization is called \emph{complete}.
Notice that the result of a complete realization is a solution of the red-black
graph if and only if such a result has no edges.
In fact, if the result \grb has at least an edge, no character is \free, by definition of realization.
Since only inactive characters can be realized, it is impossible to solve \grb.


\subsection{Reducible graphs}
In~\cite[Theorem~15]{DBLP:journals-tcs-BonizzoniBDT12} it has been proved that there exists
a relation between the existence of a phylogenetic tree solving a  red-black graph  and a special sequence of realizations, as  stated below:

\vspace{.2in}
\noindent
{\em Main Theorem:}
{\bf Given the red-black \grb instance of the PP problem, then  \grb  is solved by a tree  if and only if there exists a sequence of realizations that reduce \grb  to the empty graph}.

 \vspace{.2in}
This result motivates the following definitions:
\begin{definition}[reduction]
	\label{def:reduction}
	A \emph{reduction}  of \grb is a sequence of inactive characters whose realization (together with the removal of all red universal characters and isolated species) makes the graph empty.
\end{definition}

By the previous definition we fomally define the notion of a \emph{reducible} graph.
\begin{definition}[reducible graph]
	\label{def:reducible}
	A red-black graph $\grb$ is \emph{reducible} if and only if
	$\grb$ is  connected, has no universal and \free character, and there exists a reduction for $\grb$.
\end{definition}

In virtue of the main result proved in ~\cite[Theorem~15]{DBLP:journals-tcs-BonizzoniBDT12}, it follows that a red-blank graph that is reducible is associated to a matrix that is solved by a persistent phylogeny.

\begin{example}
	Consider a matrix $M$ having no active characters and associated with the red-black graph for the root of the tree depicted in Figure~\ref{fig:alberoesempio} and consisting  only of black edges (i.e no active character is associated with the root).
	The sequence $S= <D,C,E,A,B>$ is a reduction of the graph since after its application we obtain an empty graph.
	The figure also shows  the realization of each character in the sequence $S$ and the   red-black graphs obtained after each realization. In the figure,  a red-black graph is associated to each node $x$  of the   tree built after the realization of the sequence of characters labeling  the path from the root to the node $x$. Notice that the graph labeling each node $x$  of the tree  is the red-black graph representation of the persistent phylogeny that consists of the subtree rooted in the node $x$ and with set of active characters that are those characters that have been realized from the root of the tree to the node $x$.
\end{example}

We end the section with the following notions that will be used in the paper. Given a tree $T$ let us call \emph{initial path} of $T$ the sequence of species that ends in a node with  degree greater than two.
\begin{definition}
	\label{def:initial-species}
	Let $s_1, \ldots, s_t$ be the sequence of species labeling nodes of the initial path of a tree $T$.
	Then the species $s_2,\ldots, s_{t-1}$, if they exist, are called the internal
	species of $T$, while $s_{1}$ is called the \emph{initial} species of tree  
	$T$. 
\end{definition}
The  initial species of a tree $T$ will also be called \emph{initial} state of $T$.

We conclude this subsection by showing that given a reducible graph, then the graph without red edges that is obtained by  a  sequence of character realizations is still reducible.
This result is stated in Proposition \ref{prop:sub-reducible} that is a consequence of the following:

\begin{proposition}[subgraph induced by a reduction]
\label{prop:subset-reducible}
Let \grb be a reducible graph and $R$ a reduction of the graph. Let $R'$ be a subsequence of $R$ such that the characters in $R'$  induce  a  subgraph $G$ of \grb of only inactive characters.  Then  the sequence $R'$  is a reduction for such subgraph $G$. 
\end{proposition}
\begin{proof}
Assume that $R=<c_1, \cdots, c_n>$ is a reduction of \grb and let $<R'= c_{1i}, \cdots, c_{ik}>$ be a  subsequence of $R$. Let $G_1$ be the subgraph of \grb  induced by the characters of $R'$ such that it has only inactive characters. Indeed, by definition a reduction is a sequence of inactive characters. Let us show that $R'$ is a reduction for $G_1$. Indeed,  $R'$ is simply the restriction of $R$ to a subset of characters but the ordering is the same of $R$, where the realization of characters in the sequence $R$   leads to the empty graph. Now, when a character is realized, we remove its incident black edges and we add red edges to other species. If we do not have any character in $R \setminus R'$, then it means that we have removed  the incident edges to characters in $R \setminus R'$.  
\end{proof}

A direct application of the above Proposition is the fact that any  subgraph of a reducible red-black graph that has only inactive characters is still reducible.  

\begin{proposition}
\label{prop:sub-reducible}
Let \grb be a reducible graph and $G_1$ a connected graph obtained from \grb by a sequence of realizations of characters such that $G_1$ has only black edges.
Then $G_1$ must be reducible.
\end{proposition}

\begin{proof}
Observe that by assumption \grb is reducible. Thus    there exists a reduction $R$ of \grb and given $R'$ the subsequence of characters of $R$ that are characters of $G_1$,  then by Proposition \ref{prop:subset-reducible} $R'$ is a reduction of $G_1$, proving what required.
\end{proof}

\subsection{Tree traversals of a persistent phylogeny and reductions: a main property}

A \emph{tree traversal} of a tree $T$ is a sequence of the nodes of $T$ where
all nodes are visited exactly once and each node always precedes all of its
descendants in such a sequence~\cite{sedgewick2001algorithms}.
The following proposition links the notions of reduction on  a red-black graph with the tree traversal of a persistent phylogeny.

\begin{proposition}[Tree traversal and reductions]~\cite{Bonizzoni2016}
	\label{prop:traversal}
	Let $T $ be a solution of the red-black graph $\grb$.
	Then any sequence of positive characters labeling the nodes in a tree traversal of $T$ is a reduction of $\grb$.
	Vice versa, given a reduction of a red-black graph \grb, there exists a polynomial time algorithm that computes a persistent phylogeny $T$ solving \grb, where the reduction is a tree traversal of $T$.
\end{proposition}

Figure~\ref{fig:alberoesempio} provides an example of tree traversal and the associated reduction.
Observe that a main consequence of Proposition \ref{prop:traversal} is that each node is associated to a graph. 
In particular,  whenever a node $x$ of
the tree has more than one child $x_1, x_2, \cdots, x_k$, then a reduction of the
graph associated to $x$ consists of  the concatenation of the tree traversals of
the subtrees rooted at $x_i$, where those subtrees can be taken in any order.

Observe that each node $x_i$ will be associated to a connected graph that can be solved independently from graph associated to another node $x_j$, for $i \not= j$. 

\subsection{The Hasse diagram}

A relevant definition in the paper is that of \emph{Hasse diagram} \Hd for  a red-black graph.
\begin{definition}[Hasse diagram]
	\label{def:Hasse-diagram}
	Let \grb be a red-black graph.
	Then the Hasse diagram~\cite{ThulasiramanSwamy} (or simply diagram) $\mathcal{P}$ for
	$\grb$ is the Hasse diagram~\cite{partial-order} for the poset $(S, \leq)$  of
	all  species of $\grb$ according to the partial order where   $s_1  \leq  s_2$
	if $C(s_1) \subseteq C(s_2)$.
	If $s_1  \leq  s_2$, we also say that {\em $s_1$ is included in
			$s_2$}.\label{def:included-species}
\end{definition}
For our purposes,  the Hasse diagram of a poset $(S, \leq)$ is a directed acyclic
graph with vertices $S$ and there is an arc $(s_{i},s_{j})$ iff $s_{i} < s_{j}$
and there does not exist a species $s$ such that $s_{i}< s < s_{j}$ (see Figure~\ref{fig:poset-graph}).
Notice that it immediate to build the Hasse diagram \Hd for a red-black graph
\grb~\cite{partial-order}.

In the following, the main notions that we will use  are related to a diagram $\mathcal{P}$ and are those of \emph{source}, \emph{sink} and \emph{chain}.
A \emph{source} of a Hasse diagram \Hd is a vertex without incoming arcs, a
\emph{sink} is a vertex without outgoing arcs, and a \emph{chain} of $\mathcal{P}$   is a
direct path from a source to a sink of  $\mathcal{P}$. A \emph{chain} $<s_1, \cdots, c_k>$ is a sequence of species in containment relation, that is $C(s_1) \subset C(s_2) \cdots C(s_k)$ and for no species $s'$ it holds that $C(s_i) \subset C(s') \subset C(s_{i+1})$, for each $i$ with $1 \leq i \leq k-1$; then in the Hasse diagram  we have the arc $(s_i, s_{i+1})$ that we label  with the set of
characters $C(s_{i+1}) \setminus C(s_i) $.
A   chain is \emph{trivial} if it consists of a singleton in the diagram, while a diagram $\mathcal{P}$   consisting  only of trivial chains is called \emph{degenerate}.

\subsection{Safe species in reducible graphs}

The key notion in the construction of a solution of a   graph  is that of safe species, which builds upon the definition of species
realization (Definition~\ref{def:species-realization}).
By definition ~\ref{def:species-realization} the species must be \pendant.
Indeed, if the species includes another species, then based on the  order of realization of characters,  we can get   different graphs.

\begin{definition}[Safe species]
	\label{def:safe}
	Let \grb a reducible graph. Then a species $s$  is called \emph{safe}    if the species can be realized  in  the graph, i.e. it is \pendant and produces a graph that is still reducible.
\end{definition}

Observe that a species $s$  can be realized in a graph \grb when it does not have incident red edges in \grb or otherwise the realization of its inactive character removes the incident red edges.

\subsection{On the existence of forbidden structures}
\label{sec:forbidden-structures-skeleton}

It is interesting to note that the recognition of graphs admitting a Perfect Phylogeny, i.e. Dollo-$0$, is based on the absence of a forbidden submatrix in the input matrix.
This forbidden configuration, known as the 4-gametes~\cite{Gus91} configurations, is defined by the existence of of rows in the matrix inducing the following pairs for two given columns $c, c'$:  $(0,0)$, $(1,0)$, $(0,1)$ and $(1,1)$.
Notice that in the  Perfect Phylogeny, the root guarantees the presence  of the $(0,0)$ configuration for each pair of characters and therefore, in the graph representation, the forbidden configuration corresponds to the absence, as an induced subgraph, of the $\Sigma$-graph: the graph containing two characters $c_1, c_2$ and three species $s_1, s_2, s_3$ inducing a simple path $s_1\,c_1\,s_2\,c_2\,s_3$ on five vertices.

In the case of the \dolloone instead, the recognition of red-black graphs solved by a \dolloone tree is based on the existence of a realization of a sequence of characters  that avoids the presence of a \emph{\rs} as an induced subgraph~\cite{DBLP:journals-tcs-BonizzoniBDT12}. Given a  a red-black graph, let us call \emph{partial reduction} of the graph a realization of a set of inactive characters in the graph. 

Let us state this main property proved in ~\cite{DBLP:journals-tcs-BonizzoniBDT12}:

\begin{property}[forbidden \rs]
A red-black graph is reducible if and only if any partial reduction does not contain   a \rs as an induced subgraph.

\end{property}
This property provides a locally-testable certificate that a    partial reduction cannot be extended to obtain a tree solving the input instance, which is a key property in several of our results.

The proof that a \rs is a forbidden configuration for the persistent phylogeny
follows directly from the fact that, if $s_1\,c_1\,s_2\,c_2\,s_3$  induce a \rs
in a graph, neither $c_{1}$ nor $c_{2}$ can become \free, making impossible
their realization and hence the reduction of the \rs.
On the other hand, (black) $\Sigma$-graphs are not forbidden but entail the presence of a persistent character. The following notion is used to characterize graphs that cannot be solved by a tree.

\begin{definition}[conflicting pair]
	\label{def:conflict}
	Given two    inactive
	characters $c_1$ and $c_2$ that belong to the same connected component ${\cal G}$ of \grb and are adjacent to species $s$, then they are a \emph{conflicting pair} in  species  $s$, if
	$S(c_1) \cap  S(c_2)  \neq  \emptyset$,
	$S(c_1) \not\subseteq  S(c_2)$ and  $S(c_2) \not\subseteq  S(c_1)$ and the connected component ${\cal G}$ has a  species $s'$ with $s'\notin S(c_1) \cup S(c_2)$.
\end{definition}

Notice that a conflicting pair induces the 4-gamete configuration.  
Therefore,  as stated in the following Lemma,   given a species  $s$ in the graph having a conflicting pair, the realization of $s$ is possible only after the removal in the graph of species that induce the conflicting pair.

\begin{lemma}
    \label{lem:conflict-pair}
    Let $s$ be a species that has a conflicting pair $(c_1,c_2)$  in a graph $\grb$. Then the realization of $s$ leads to a red $\Sigma$-graph. 
    \end{lemma}

    \begin{proof}
    Observe that $s$ in the graph must be \pendant for the species being realized,  i.e. there does not exist a species having only character $c_1$  or $c_2$ that is included in $s$. By definition of conflicting pair there exists at least a species $s_1$ with $c_1$ and some character that is not in $s$ and similarly there exists a species $s_2$ with $c_2$ and some character   not in $s$, i.e. $s_1$ and $s_2$ cannot be removed by the realization of $s$. This fact implies that any order of realization of the characters in $s$ does not remove neither $s_1$ nor $s_2$ and thus we obtain a \rs in the graph.
\end{proof}


In the paper we will extensively use Lemma \ref{lem:conflict-pair} in the proofs of many technical Lemmas.

The previous discussion leads us to the natural question if it is possible to decide if a graph can be solved by a tree by using a characterization in terms of forbidden substructures, similarly to the case of the perfect phylogeny.
The problem seems not trivial, even in the case of maximal characters addressed in this paper. Indeed, it has been proved the existence of an infinite family of forbidden induced subgraphs in this restricted case~\cite{bonizzoni_when_2014}.

\section{Organization of the main results}
\label{sec:paper-structure}

First notice that we can restrict ourselves to reducible red-black graphs.
In fact, a red-black graph \grb is not reducible if (1) is disconnected, (2) has
a universal character, (3) has a \free character, or (4) does not have a solution.
In case (1) we can solve each connected component separately, then a
concatenation of the reductions is a solution of \grb, in case (2) we can
realize the universal character, then solve the resulting red-black graph, in
case (3) we can remove the \free character, and to manage case (4) it suffices
to check if the reduction obtained is a actually a solution of \grb.
{\bf Therefore, in the remainder of the paper we can always assume that \grb is reducible.}

Thus, unless specified differently the results presented in the next sections implicitly assume that the red-black graph is reducible.

The paper has two parts: the first part is dedicated to solving
red-black graphs whose characters are all maximal or active, while the second part extends that result to
general red-black graphs.

The connection between the two parts is due to the fact that, in a tree solving
a reducible red-black graph, the edges incident on the root are labeled by
maximal characters (Proposition~\ref{prop:maximal-first}): the subgraph induced by such characters
will be called \emph{skeleton}.
Moreover, given any solution $T$, removing all labels that do not correspond to
maximal characters and contracting all unlabeled edges results in a
solution of a skeleton.
Those two properties mean that an algorithm to solve a red-black graph \grb is to
solve the skeleton of \grb, obtained by deleting all characters that are
not maximal, then realizing the first character (that is, the
character labeling an edge incident on the root) of the solution of the
skeleton, and finally iterate on the red-black graph obtained after the realization.

The above result is discussed in section \ref{sec:an-invariant-propery}
\emph{An invariant property for red-black graphs reductions}.

It is immediate that this procedure is correct if the skeleton has only
one solution, i.e. a unique tree that solves the skeleton,  since in that case there is no ambiguity on the character to realize.
If the skeleton has more solutions,  then we show that we need to iterate the realization of a safe species (see definition \ref{def:safe}) and such species can be computed in polynomial time \ref{lem:polynomial}.

We are able to show  that there exists a unique tree solving a skeleton and starting with a given safe species (see section \ref{sec:characterization}). This fact will implies that after the realization of a safe species for the skeleton, the general graph admits only one possible reduction based on the choice of this initial species.

Now, in order to find  a safe species that is also safe for the general graph, we iterate the realization of  a safe species   for the skeleton,  till we get a reducible graph without red edges, on which to iterate again the  procedure till we get the reduction of the whole graph.
 Observe that, based on Proposition \ref{prop:sub-reducible},  whenever a solvable graph is reduced to a graph with no red edges, then it must be solvable too. 
 We will show that the number of possible solutions for a skeleton graph is at most the number $m$ of characters and in most cases is bounded by $2$.
Moreover, we show that when a  safe species of the skeleton is realized, then  there exists a unique  reduction that corresponds to the unique tree that has that given safe species. Since after the realization of a safe species   the reduction is unique, it follows that  we iterate this step at most    $m$ times. Since  each iteration costs  polynomial time in the size $n \times m $ of the graph,  we are able to guarantee a polynomial time solution.

In Section \ref{sec:the structure-skeleton} we will  show that some skeletons, called
\emph{degenerate}, can admit more than two solutions (\ref{thm:skeleton with two safe species}). Moreover,  degenerate skeletons are trivial to solve (\Cref{prop:degenerate}).
Instead, if the skeleton is not degenerate, then it is possible to iterate the realization of a safe species (see definition \ref{def:safe}) and such species can be computed in polynomial time \ref{lem:polynomial}.

To solve skeletons that are not degenerate we analyze the Hasse diagrams
associated with a skeleton, focusing on finding a vertex (i.e. the safe species,
\Cref{def:safe}) of the Hasse diagram that determines the first characters to
realize.
To find a safe species in polynomial time (\Cref{lem:polynomial}) we will need
to prove some properties of chains of Hasse diagrams.
In particular, the time complexity is due to a polynomial time characterization of chains having safe species.

Part II of the paper is focused on presenting the main technical results that lead to the polynomial time algorithm to solve general graphs.

More precisely, the polynomial  algorithm for the general graphs iterates the computation of the safe species of the skeleton that keeps invariant the property of the skeleton graph of being reducible.  
This is Algorithm \ref{alg:PPR} that applies the iterative procedure that we have discussed above and relais  on the fact that after the first characters of a solvable graph have been realized, then the reduction is unique.

\section{Maximal characters and skeleton graphs}
\label{sec:m-characters}

As outlined in Section~\ref{sec:paper-structure}, in the first part of the paper we
focus on maximal characters and on graphs that only have
maximal characters, called \emph{skeleton graphs}, since
in a tree solving a reducible red-black graph \grb, the edge incident on the
root is labeled by maximal characters (\Cref{prop:maximal-first}).
Let us first introduce the definition of maximal character.

\begin{definition}[maximal character]
	\label{def:maximal-character}
	Let \grb be a red-black graph, and let $c$ be an inactive character of \grb.
	Then $c$ is \emph{maximal} iff there is no inactive character $c_{1}$ with $S(c)\subset S(c_{1})$. 
	We denote with $C_M$ the set of inactive maximal characters of \grb.
\end{definition}

\begin{definition}[Comparable characters]
	\label{def:comparable-characters}
	Let \grb be a red-black graph, and let $c_1$, $c_2$ be two characters of \grb.
	Then $c_1$ and $c_2$ are \emph{comparable} if $S(c_1)\subseteq S(c_2)$ or $S(c_2)\subseteq S(c_1)$.
	Two characters that are not comparable are called \emph{incomparable}.
\end{definition}

Notice that the term maximal character refers to inactive characters, while the notion of comparable and incomparable characters refer also to active characters.
Moreover, two maximal characters are either identical or incomparable.
The following notion of \emph{skeleton} graph is crucial in our approach to solve the \dolloone problem.

\begin{definition}[skeleton graph]
	\label{def:skeleton graph}
	Let $\grb$ be a red-black graph.
	The \emph{skeleton graph},  in short \emph{skeleton},  of \grb is the subgraph  of \grb denoted as \grbm which is  induced by the set of vertices $C_M  \cup N(C_M)$,  that is the subgraph induced by the maximal characters and its adjacent species.
	If two characters  of $C_{M}$ have the same set of species, only one of them is in \grbm (it is indifferent which one).
\end{definition}

The \emph{active skeleton} is obtained similary, but keeping all maximal  and all  active characters.
In particular, given a skeleton graph, with the realization of inactive characters we may obtain active characters, that is an active skeleton graph.

\begin{definition}[active skeleton graph]
	\label{def:extended-skeleton graph}
	Let $\grb$ be a red-black graph and let $\A$ be the set of its active characters.
	The \emph{active skeleton graph} \grbe of \grb is the subgraph induced by the set of vertices $C_M \cup \A \cup N(C_M \cup \A)$, that is, the subgraph induced by the maximal characters, the active characters, and their adjacent species.
	If two characters of $\grbe$ have the same set of species, only one of them is in \grbe (it is indifferent which one).
\end{definition}

In particular, we call \emph{skeleton tree}  of a tree $T$, the tree that is induced in $T$ only by maximal characters.
More precisely, the skeleton tree of tree $T$ is obtained from tree $T$ by contracting  all edges labeled only by non maximal characters.

Observe that the skeleton graph is clearly a subgraph of the active skeleton
graph.
A similar relation also holds on \emph{solutions} of the PP problem over red-black graphs and it is the basis of our
algorithm.
Given a red-black graph \grb and a solution $T$ of \grb, there is a subtree
$T_{E}$ of $T$ that is a solution of the active skeleton \grbe of \grb, and
there is a subtree
$T_{M}$ of $T_{E}$ that is a solution of the skeleton \grbm of \grbe (and of
\grb), where such subtrees are obtained by contracting edges of $T$.

\begin{figure}[tb!]
	\begin{subfigure}{.475\linewidth}
		\includegraphics[width=0.5\linewidth]{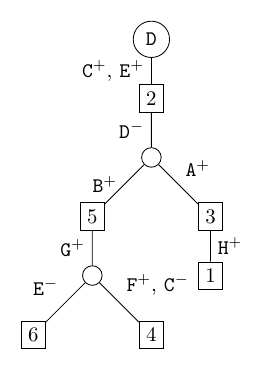}
		\caption{Solution of the red-black graph \grb on the right.
			The root is labeled with the active characters.}
		\label{fig:skeletons:te}
	\end{subfigure}\hfill 
	\begin{subfigure}{.475\linewidth}
		\includegraphics[width=\linewidth]{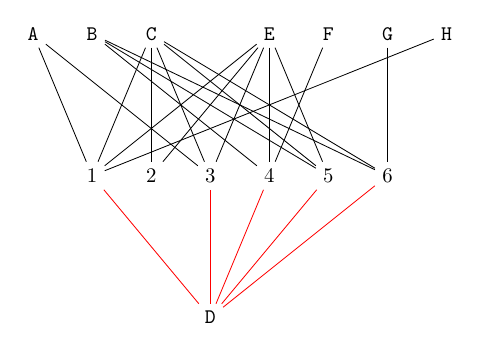}
		\caption{Red-black graph \grb}
		\label{fig:skeletons:grb1}
	\end{subfigure}

	\medskip 
	\begin{subfigure}{.475\linewidth}
		\includegraphics[width=0.4\linewidth]{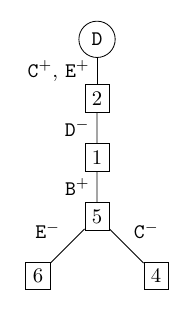}
		\caption{Subtree $T_{E}$ of $T$ solving the active skeleton \grbe on the right}
		\label{fig:skeletons:t}
	\end{subfigure}\hfill 
	\begin{subfigure}{.4\linewidth}
		\includegraphics[width=\linewidth]{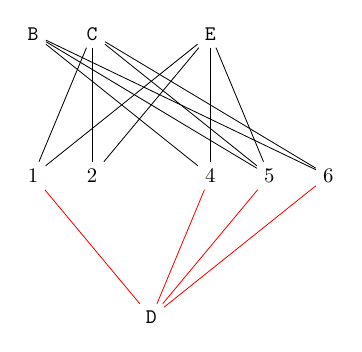}
		\caption{Active skeleton \grbe of the red-black graph \grb}
		\label{fig:skeletons:grb2}
	\end{subfigure}

	\medskip 
	\begin{subfigure}{.475\linewidth}
		\includegraphics[width=0.5\linewidth]{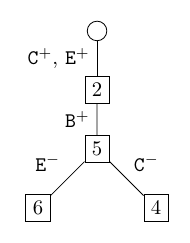}
		\caption{Subtree $T_{M}$ of $T_{E}$ solving the  skeleton \grbm on the right}
		\label{fig:skeletons:tm}
	\end{subfigure}\hfill 
	\begin{subfigure}{.475\linewidth}
		\includegraphics[width=\linewidth]{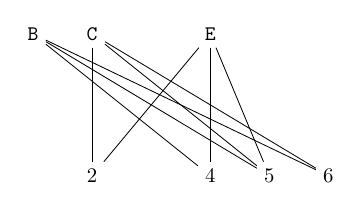}
		\caption{Skeleton \grbe of the red-black graph \grb}
	\end{subfigure}

	\caption{Example of red-black graph, its active skeleton and skeleton, and
		their respective solutions.}
	\label{fig:skeletons-example}
\end{figure}

\subsection{An invariant property for red-black graphs reductions}
\label{sec:an-invariant-propery}

The  following  proposition provides a motivation of our interest in finding trees solving skeleton graphs:
we can always assume that the first character in a reduction of a general red-black graph is a maximal character of a reduction for a skeleton graph  (see  Proposition~\ref{prop:maximal-first}).
For a clearer discussion of the results and preliminaries,  all the proofs in the section are presented in a separate section.



\begin{proposition}
	\label{prop:maximal-first}
	Let \grb be a reducible red-black graph and let \grbM be its skeleton graph.
	Then there exists a reduction $R$ of $\grb$ that starts with a maximal inactive character $c$.
	Moreover, there exists  a reduction of \grbM that also starts with $c$.
\end{proposition}

Observe that  Proposition~\ref{prop:maximal-first} holds for general graphs.  Since it states an invariant property of general reducible graphs, it induces a recursive schema for solving  general reducible graphs.

We leave the solution of the general step as the Part II of the paper.

The following lemma is necessary since the skeleton of a reducible red-black
graph is not necessarily connected.

\begin{lemma}[the skeleton and the active skeleton are reducible]
	\label{lem:da-fare}
	Let \grb be a reducible graph. Then  \grbe and each connected component of  
	\grbM is a reducible graph.

\end{lemma}

Proposition \ref{prop:maximal-first} allows us to establish the following key property:

\begin{remark}
\label{re:invariant-maximal}
{\bf Main Invariant Property} Any tree $T$ solving a reducible red-black graph  starts with a maximal inactive character which is also the topmost character in a tree solving its skeleton graph.
\end{remark}

However notice  that  a skeleton graph $\grbm$ could be solved by more than one tree, and some of the potential trees solving $\grbm$ could not be extended to include minimal characters, as shown in the following example.

\begin{figure}[hbt]
	\centering
	\includegraphics[scale=0.6]{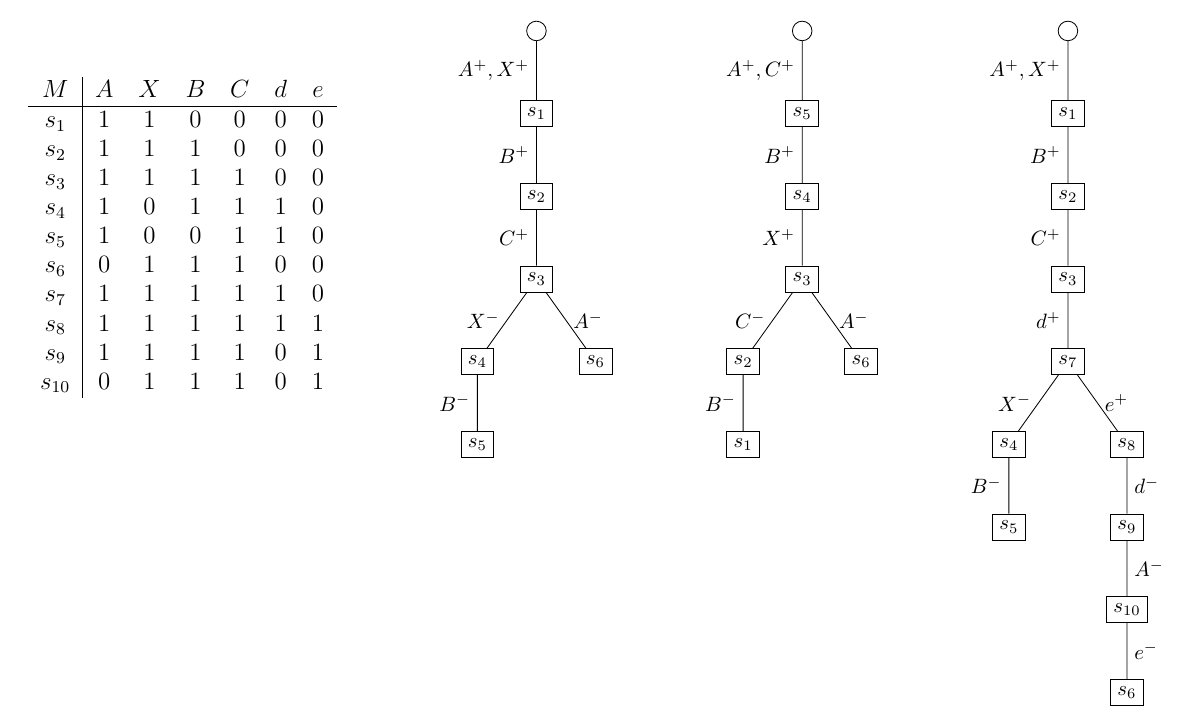}
	\caption{An instance of the PP problem (left) having two possible reductions for the skeleton graph (center), and the tree $T$ solving the input matrix (right).
		Only the first solution can be extended to a tree solving the complete instance.
		Conversely, the tree induced by maximal characters in $T$ corresponds exactly to the first solution.}
	\label{fig:maximal-minimal}
\end{figure}

\begin{example}
	\label{ex:2skel}
	The example in Figure~\ref{fig:maximal-minimal} contains ten species and six characters, two of which are not maximal ($d$ and $e$).
	In its skeleton graph, the species within the sets $\{s_3,s_7,s_8,s_9\}$ and $\{s_6,s_{10}\}$ are equivalent since they share their set of maximal characters.
	The skeleton graph admits two reductions, one starting with the realization of species $s_1$ and the other starting with $s_5$.
	Nevertheless, only the first one admits the insertion of not maximal characters $d$ and $e$ to extend the tree to a solution of the original input matrix.
	Indeed, observe that $A$ and $e$ are conflicting characters and, since $A$ is the first character to be realized, it is necessary to realize all species that have character $A$ and do not have character $e$.
	In the second solution, this can not be done since the realization of $s_5$ forces the realization of $d$ and $C$, preventing the realization of species $s_1$ and $s_2$ and inducing a \rs.
\end{example}

\section{The  structure of  a tree solving a skeleton graph}
\label{sec:the structure-skeleton}

In this section, we show that trees solving
skeleton graphs have a well defined structure.

Given a tree $T$ and two edges $e_1$ and $e_2$, we say that $e_1$ is \emph{above} $e_2$ (and $e_2$ is \emph{below} $e_1$) if $e_1$ lies on the path of $T$ starting at the root and ending with the edge $e_2$.
The following proposition
for skeleton graphs, describes the possible relations between the positive and
negative labels of maximal inactive characters in a tree solving an active skeleton graph.
 
\begin{proposition}[ordering of negated characters]
	\label{prop:basic-1}
	Let \grbe be an active skeleton graph, and let $T$ be a tree solving $\grbe$.
	Let  $c_1$ and $c_2$ be two maximal inactive characters of $\grbe$. Then:
	\begin{enumerate}
		\item $c_1^-$ and $c_2^-$ label two distinct edges, if both exist;
		\item if $c_1^+$ and $c_2^+$  label the same edge of $T$, then the edges labeled by $c_1^-$ and $c_2^-$ exist and none of them is below the other;
		\item if the edge labeled by $c_1^+$ is above the edge labeled by $c_2^+$, then the edge labeled by $c_1^-$ is not below the edge labeled by $c_2^-$ (if the latter exists), and if $S(c_1)\cap S(c_2)\neq\emptyset$ then the edge labeled by $c_1^-$ is below the one labeled by $c_2^+$.
	\end{enumerate}
\end{proposition}

\begin{proof}
	
	Let us first show that statement 1) holds. Clearly, if  $c_1^-$ and $c_2^-$ label the same edge, by definition of tree solving \grbe it must be that $c_1^+$ and $c_2^+$  both occur along the same path before the edge labeled $c_1^-$ and $c_2^-$. Since by definition $c_1^+$ and $c_2^+$ are maximal characters, then either they are not comparable or they are disjoint characters. This last case is not possible since $c_1^+$ and $c_2^+$  both occur along the same path before the negation of both characters. But, if they are not comparable, one positive character occurs before the other one which implies that the character above the other has the  species of the other character, as they are negated on the same edge. So $c_1^-$ and $c_2^-$ cannot label the same edge  since this fact contradicts the maximality of one of the two characters. Thus statement 1) follows.
	Let us now show statement 2). This is a consequence of the fact that  if edges labeled by $c_1^-$ and $c_2^-$ exist and one is below the other, for example $c_1^-$, it follows that the the set of species of $c_1$ include those of $c_2$, thus contradicting the maximality of $c_2$.
	Finally, let us prove the statement 3) of the Proposition. Again w.l.o.g. assume that $c_{1}^{+}$ precedes $c_{2}^{+}$ in $T$ (the other case leads to symmetric conclusions). If an edge labeled $c_{2}^{-}$ exists, then it cannot precede $c_{1}^{-}$,
	otherwise the character $c_{2}$ is contained in $c_{1}$.
	By the maximality of $c_{2}$, in the case that  $c_{1}$ and $c_{2}$ are overlapping, i.e. they share a common species, then  $c_{2}^{+}$ must precede $c_{1}^{-}$ in $T$, otherwise they are disjoint characters.

\end{proof}

\begin{figure}[bt]
	\centering
	\includegraphics[width=6cm]{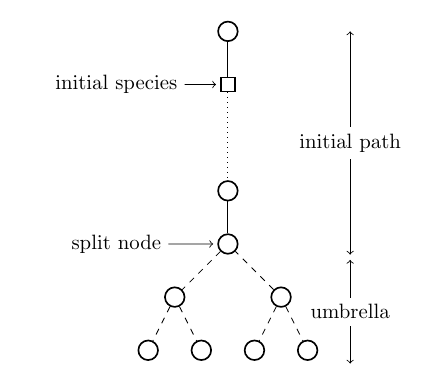}
	\caption{Structure of a tree solving a set of maximal characters, dashed edges contain only negative labels.
		The tree is composed by an initial path containing all character gains, the initial path finishes in a split-node under which only character losses are allowed.}
	\label{fig:tree_maximal}
\end{figure}

The following Lemma describes the structure of a tree solving a skeleton graph.

\begin{lemma} \label{lemma:skeleton-shape}
	Let   $T$ be a tree solving a connected skeleton \grbM on a set $C_{M}$ of
	maximal characters.
	Then  $T$ is composed of at most two parts sharing a single node
	(Figure~\ref{fig:tree_maximal}): (1) a simple path starting from the root,
	called \emph{initial path}, where all characters of $C_M$ are gained (and
	some characters might be negated) and, optionally, (2)
	a tree, called \emph{umbrella},  whose root has at least two children and such that all of its edges are labeled by character losses.
\end{lemma}

\begin{proof}
	If the tree $T$ solving the graph \grbM is a path, then the result is
	trivially true since it corresponds to the case that the umbrella does not exists.

	Let us assume that $T$ is not a simple path, and let $z_1$ be the vertex of $T$ that has
	at least two children and is closest to the root. We need to prove that no
	character gain can happen below $z_1$.

	Since \grbM is reducible and connected, it has no \free character, the root has only one
	child $x$ and the edge incident on the root must be labeled only by positive characters.

	Assume that there exist two (maximal) characters $c_1$
	and $c_2$ that are gained in two separate branches of $T$, and assume that
	those are the two characters with such property that are closest to the
	root.
	Let $d^+$ be any positive
	character that is above both $c_1^+$ and $c_2^+$
	--- at least one such character exists by our previous argument.
	Moreover, let $z$ be the lowest common ancestor of the edges labeled by
	$c_1^+$ and $c_2^+$: since $z$ has at least two children while the root has
	only a child, then $z$ cannot be the root of $T$.

Let us first consider the case that there exists a common species that is shared by $d$, $c_1$ and a common species that is shared by $d$, $c_2$,  then
	Proposition~\ref{prop:basic-1}, statement 3)  implies that $d^-$ must be below both $c_1^+$
	and $c_2^+$ which is impossible. Thus this case is not possible.

Consider now the case that $d$ and $c_1$ share no common species; observe that this case is symmetric to one where $d$ and $c_2$ share no common species.  Since $d^{+}$ is above
	$c_1^{+}$, this can only happen if $d^-$ is above  the occurrence of the species where $c_1^+$ is gained.
	If $d^-$ is below $z$, then $c_2$ is contained in $d$, contradicting the
	maximality of $c_2$. Indeed, being  $d^-$ before the occurrence of   $c_1^+$ it follows that $d^-$ cannot be before and after the occurrence of  $c_2$. Consequenly,  $d^-$ is above $z$.
	By the generality of $d$, all characters that are gained above
	$z$ are also lost above $z$.

	Consequently, the state of $z$ is equal to the state of the root of $T$. This fact implies that   \grbM cannot be connected, which contradicts the initial assumption.
	Therefore there cannot be two characters gained on two
	separate branches.

	We have not ruled out yet that a single character $c$ is gained below the split
	node $z_1$.
	Since $z_1$ has at least two children, there is a branch stemming from the split-node
	that does not contain the edge $c^{+}$.
	By the above argument (i.e. that two positive characters are gained on two distinct branches below the split-node), such a branch cannot contain any character gain; let
	$e^{-}$ be the label of the edge incident on $z_1$ of such a branch.
	Clearly, $e^{+}$ is above $z_1$.
	But this implies that $S(c)$ is included in $S(e)$, contradicting the
	assumption that $c$ is maximal, concluding the proof that the umbrella can
	only contain negative characters.
\end{proof}

If the umbrella does not exist, the tree $T$ is called a {\em line-tree}.
Otherwise, the tree $T$ is called a \emph{branch-tree} and the root of the
umbrella is called the \emph{split-node}.

In particular, we call \emph{pseudo-line} tree a branch-tree that has an
umbrella with only two branches, one that has only one edge and such edge is labeled by the negation of the last character that has been introduced before the split-node. 
Notice that, in this case the graph can be solved also by a line-tree.

\begin{figure}[htb!]
	\centering
	\includegraphics[height=.33\textheight]{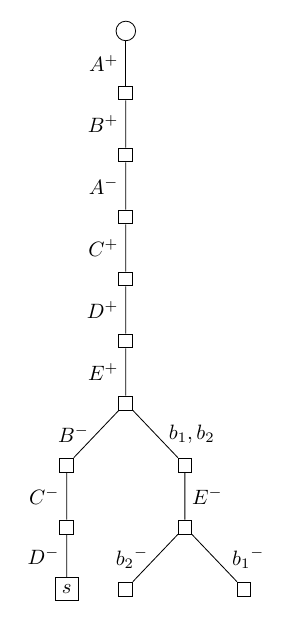}
	\caption{An example of tree such that its skeleton is  a pseudo-line tree solving a skeleton graph. The last gained character $E$ is negated just below the split-node of the tree. 
 }
	\label{fig:pseudo-line}
\end{figure}


\begin{lemma}
	\label{lem:all C gained in initial path}
	Let $T$ be a tree solving an active skeleton graph $\grbe$.
	Then all inactive characters of \grbe are gained in the initial path of $T$.
\end{lemma}

\begin{proof}
	Assume to the contrary that a character $c$ is gained in the umbrella of $T$,
	that is below the split-node $z$.
	If $z$ has any character $d$ of \grbm, then $d$ includes $c$, contradicting the
	hypothesis that \grbm is a skeleton graph.
	Hence $z$ has no character, but this contradicts the assumption that \grbm is
	connected, completing the proof.
\end{proof}

Notice that a solution of a red-black graph might not have an initial species.

The following property is a direct consequence of Proposition~\ref{prop:basic-1}, statement 2 by which if two characters label the same edge, then their negation cannot lay in the same path.

\begin{corollary} \label{cor:initial-2-c}
	Let $T$ be a tree solving a connected skeleton graph $\grbM$.
	If  the initial species of a tree $T$ solving $\grbM$ has at least two inactive maximal characters,
	then no maximal character is negated along the initial path of tree $T$.
\end{corollary}

In fact, all inactive characters of the initial species must be lost in
different branches.
Moreover, no character gained below the initial species can be lost in the
initial path, otherwise it is included in a character of the initial species and thus the tree cannot have only maximal characters.

\subsection{Degenerate skeletons}

\begin{figure}[htb!]
	\centering
	\includegraphics[width=4cm]{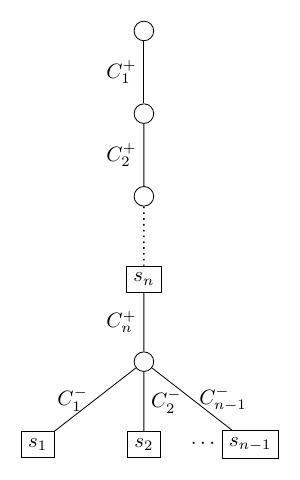}
	\caption{A tree solving a degenerate graph consisting of the characters $C_1, \ldots, C_{n}$. The species $s_i$ contains all characters but the character $C_i$. The tree has an initial state given by the species $s_n$ with characters $C_1, \ldots, C_{n-1}$.
		An alternative tree solving a degenerate graph will have also the species consisting of all characters. }
	\label{fig:degenerate}
\end{figure}

A special class of skeleton graphs is the family of \degenerate skeletons, where
for each maximal character $c$ there is a species that has all maximal characters except $c$.

\begin{definition}[\degenerate]
	\label{def:degenerate}
	A red-black graph \grb with inactive characters $I$, is called
	\emph{\degenerate} if there exists an injective function $\phi: I\mapsto S$ such that
	for each inactive character  $c_i\in I$, the inactive characters of the species $\phi(c_i)$ is the set $I\setminus \{c_i\}$.
	Moreover, $S$ consists exactly of the set $\{\phi(c_i) : c_i\in I\}$ and eventually a species $s$ whose set of inactive characters is exactly $I$.
\end{definition}

A  \degenerate graph may have different solutions,  all consisting of an initial
path where all characters are gained and no character is lost, while
all  characters, except for one,  are lost in a distinct edge of the umbrella, incident on
the split-node.
The only character that is not lost in the umbrella is gained in the edge of the
initial path incident on the split-node.

Notice that all possible solutions differ in the species that label nodes of
the initial path and in the character labeling the last edge of the initial path.
More precisely, the initial path consists of a permutation of all inactive
characters except for a character $c_i$, the character $c_{i}$ is gained in the
last edge of the initial path, and all other character are lost in the umbrella
(see Figure~\ref{fig:degenerate}).
Since \degenerate red-black graphs can be recognized in polynomial time and all
its solutions can be generated in polynomial time, in the following we can focus
on reducible graphs that are not degenerate.


\begin{proposition}
	\label{prop:all-degenerate-solutions}
	Let \grb be a \degenerate skeleton red-black graph, let $T$ be a solution of \grb, and
	let $c^{+}$ be the label of the last edge of the initial path of $T$.
	Then all leaves of the umbrella of $T$ are adjacent to the split-node, and all
	characters different from $c$ are lost in the umbrella.
\end{proposition}

\begin{proof}
	A consequence of Lemma~\ref{lemma:skeleton-shape} and the definition of
	\degenerate graphs.
\end{proof}

\subsubsection{Existence of an initial species in non degenerate skeletons}

Now that we have a characterization of the solutions of a \degenerate red-black
graph, we can prove that all other skeleton trees must have an initial species,
that is a species labeling a node of the initial path.

\begin{lemma}
	\label{lem:initial-species-exists}
	Let \grbe be an active skeleton, and let $T$ be a solution of \grbe.
	If $T$ has no initial species, then \grbe is   \degenerate.
\end{lemma}

\begin{proof}
	Let $T$ be a solution of \grbe that has no initial species.
	Clearly, $T$ cannot be a line-tree, since all species lie on the initial
	path of any line-tree.
	By \ref{lem:all C gained in initial path}, the split-node of $T$ has all
	inactive character of \grbm.
	By Proposition \ref{prop:basic-1}, all inactive characters of \grbm are lost in different
	branches, below the split-node.
	It is immediate to build the injective function that proves that \grbe is \degenerate.
\end{proof}

The following corollary is the main ingredient of our procedure to compute the
first character of a reduction.

\begin{corollary}
	\label{cor:initial species guaranteed}
    Let \grbe be an active skeleton that is not \degenerate, and let $T$ be a tree in standard form that solves \grbe.
    Then $T$ has an initial species.
\end{corollary}

Thus we conclude the section stating the following fact.
\begin{proposition}
	\label{prop:degenerate}
	Let \grb be a red-black graph.
	Then we can determine in polynomial time if \grb is \degenerate and, in case, we
	can enumerate all its solutions in polynomial time.
\end{proposition}

\section{The polynomial time algorithm for solving skeleton graphs}
\label{sec:skeleton-graphs}

Let us first give an outline of the main steps that lead to a polynomial time algorithm for solving a skeleton graph.
For convenience all the proofs are in a separate section.
Observe that after the realization of a species a skeleton graph can become an active skeleton graph, as some characters may be active.
The main idea of the algorithm is to have a sufficient condition for having a safe species for an active skeleton, where a safe species is defined in definition \ref{def:safe} as the one whose realization preserves the property of the graph of being reducible. 
Observe that a safe species for a skeleton graph may not be safe for the active skeleton.

\begin{observation}
	We first show that a reduction of a non \degenerate active skeleton always starts with the realization of a \emph{safe} species: by definition  such species keeps invariant the property of having components that are reducible, after their realization.
\end{observation}

\begin{observation}
	We provide a polynomial time algorithm for computing safe species for active skeleton graphs. The algorithm is based on the construction of the Hasse diagram for the associated  skeleton graph. More precisely, we show that {\bf  a necessary and sufficient condition for a species $s$ to be safe for $\grbe$ is that $s$ is the source of an \emph{active chain} of the Hasse diagram (\Cref{cor:property-necessary-sufficient}) for the associated skeleton graph.}
\end{observation}

\begin{observation}
	Assuming that the graph is reducible and is not \degenerate{}, the iterative process of realizing a safe species results in a reduction of the graph. In particular, either the process stops finding a \degenerate graph in which case Proposition \ref{prop:degenerate} applies or it stops by having the empty graph, i.e. a reduction is obtained,  or  if the graph is not reducible, then  the iteration process does not find any  safe species  in the graph, which corresponds to the case that the graph has an induced red $\Sigma$-graph and hence no solution exists.

\end{observation}

\subsection{Safe species in (active) skeletons}


In this section we show how to compute safe species in  active skeleton graphs.
The computation of safe species in an active skeleton graph is based on the Hasse diagram for the associated skeleton graphs and in particular on chains of the Hasse diagram whose realization in the active skeleton graphs does not induce the forbidden red $\Sigma$-graph. Indeed, we will show that the source of such chains is a safe species of the active skeleton and this condition is not only necessary but sufficient.

\begin{figure}[ht]
  \centering
\includegraphics[width=9cm]{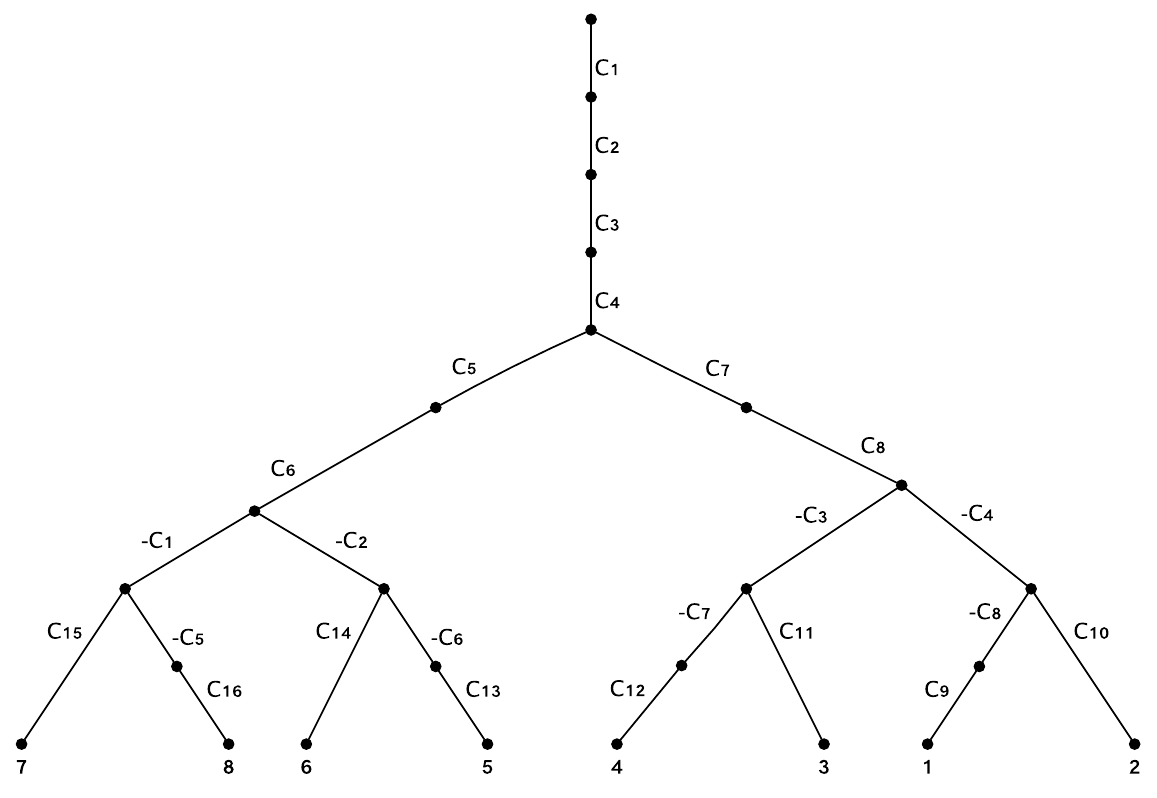}
\caption{An example of  tree that has minimal and maximal characters and such that all species are on the leaves (the only species are indicated by the numbers) and there exists no species that can be realized in the associated red-black graph. Indeed, observe that the tree is built from an initial-path consisting by the realization of the maximal characters $c_1, c_2, c_3, c_4$ and all the remaining characters are minimal ones. The skeleton of the tree is the degenerate tree. 
}
\label{fig:albero1} 
\end{figure}

The existance of a  safe  species is only
garanteed in (active) skeleton graphs that are not \degenerate as stated in
Proposition~\ref{prop:initial-species}. Indeed a general graph may be solved by a branch-tree having no species that can be realized in the graph. Figure \ref{fig:albero1} illustrates an example of  tree solving a  graph that does not have an initial species. Observe that the skeleton of such graph is a degenerate one.

\begin{proposition}
	\label{prop:initial-species}
	Let $\grbM$ be a skeleton graph that is not \degenerate.  Then for any tree $T$ solving the graph,   $T$ has an initial species $s$ that is safe for $\grbM$.
\end{proposition}
\begin{proof}
By Lemma \ref{lem:initial-species-exists} it must be that $T$ has an initial species $s$.
Then   $s$ is safe because of the bijection between tree $T$ and a reduction of graph \grbM. Indeed, the realization of $s$ applied to  \grbM produces a  graph  that is solved by the subtree $T_s$ with root $s$. By definition $s$ is safe for graph  \grbM if its realization applied to graph \grbM produces a reducible graph.
\end{proof}

The above proof can be extended to show the same result for the active skeleton graph of a general graph.
\begin{proposition}
	\label{prop:initial-species-extended}
	Let $\grbe$ be the active  skeleton graph for a graph \grb such that  its \grbM is not \degenerate.  Then the subtree of  tree $T$ solving \grb that is a solution of the active skeleton graph  has an initial species $s$ that is safe for $\grbe$.
\end{proposition}


\subsection{The Hasse diagram: how to compute safe species}

The following definitions are referred to general  red-black graphs, more precisely we define the  \emph{realization}  of 
a \emph{chain} $\mathcal{C}$  of the Hasse diagram   for the skeleton graph  of a graph \grb.

\begin{definition}[realization of a chain]
	Let $\mathcal{C} = <s_1, s_2, \ldots, s_k>$ be a chain of the diagram
	$\mathcal{P}$.
	Then the  \emph{realization} of  $\mathcal{C}$ in a graph is the
	realization in the graph of characters of species
	$s_{1}$
	followed by the realization of characters labeling the arcs $(s_{i}, s_{i+1})$ for $1\le i\le k-1$, in
	any order.
\end{definition}

\begin{remark}
	In particular we will be interested in finding active chains for skeleton graphs, as defined below.
	Let us recall that we need to consider active characters in \grbe to find those chains of \grbm that are of interest for the invariant property stated in Proposition \ref{prop:maximal-first}: we need to find the initial characters of reductions  solving active skeleton graphs that are also initial in reductions   solving general graphs.

\end{remark}

Le us give the following crucial definition of \emph{active chain} that will allow us to characterize safe species for active skeleton graphs.

\begin{definition}[active chain for \grbe]
	\label{def:active-chain}
	Let $\grbe$ be an active skeleton and \grbm its skeleton graph.
	Let $\cal C$ be a nontrivial chain of \grbm.
	Then $\cal C$   is  \emph{active} in  \grbe  if its source $s$ can be realized   in \grbe
	and the realization of the chain in the graph \grbe  does not induce any red $\Sigma$-graph.

\end{definition}

Observe that the above definition of active chain imples that the source of the chain must have all active characters of the graph \grbe or the realization of the characters of the source of the chain will remove the active characters from the graph.

\begin{figure}[tb!]
	\centering
	\includegraphics[width=\textwidth]{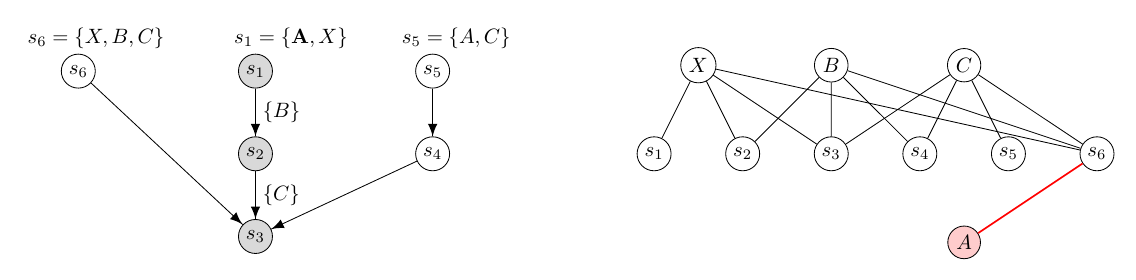}
	\caption{Example of Hasse diagram with an active chain.
		On the left is represented the Hasse diagram of the active skeleton
	red-black graph that is on the right, where \texttt{A} is the only
		active character of the graph.
		The chain $<s_1, s_2, s_3>$ (with gray background) is active, since the source $s_{1}$  has all
		active characters of \grb.
	Now $s_1$ is a safe species since the realization of the chain does not induce red $\Sigma$-graphs; indeed Figure~\ref{fig:poset-tree} represents the tree solving the graph.}
	\label{fig:poset-graph2}
\end{figure}

The following notions will be used to prove the necessary and sufficient condition for having safe species.

\begin{definition}[complete chain]
	\label{def:simple-complete}
	Let $\cal C$ be a chain of the Hasse diagram for a reducible graph $\grb$.
	Then $\cal C$ is {\em complete} if its sink has all inactive characters of $\grb$.
\end{definition}

If a chain is not complete, it is called {\em incomplete}.
Similarly, we  define the notion of complete/incomplete trees and graphs.

\begin{definition}[complete/incomplete trees and graphs]
\label{def:complete grb}
An active skeleton graph  is \simple if it solved by a tree  $T$   having  a species that includes all inactive characters of $T$. Then the tree $T$ is  called \simple and in all other cases  $T$ and the associated skeleton graph is {\em \incomplete}.

\end{definition}

The following Proposition is an immediate consequence of the definition of \simple tree and 
graph and that each species is trivially  included in the species that has all characters.

\begin{proposition}\label{lem:property-branch-complete}
	Let $\grbe$ be a  \simple skeleton graph.
	Then, all  chains of $\grbe$ are complete.
\end{proposition}


\begin{figure}[htb!]
	\centering
	\includegraphics[width=0.9\textwidth]{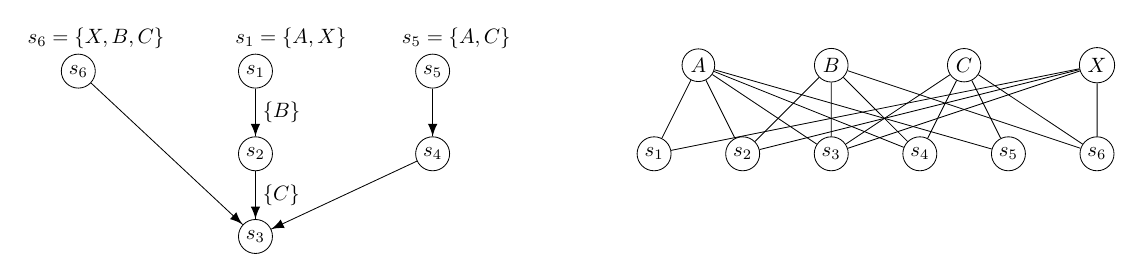}
	\caption{Example of Hasse diagram and the associated graph.  In the chain $\mathcal{C} = <s_1, s_2, s_3>$, we have that $s_1 = \{A,X\}, s_2 = \{A,X,B\}$, and $s_3 = \{A, X, B, C\}$.
	Moreover, $\{B\}$ labels the arc $(s_{1}, s_{2})$  and $\{C\}$ labels the arc $(s_{2}, s_{3})$, then the sequence $<A,X,B,C>$ is a realization of chain $\mathcal{C} $. Observe that the order of characters of each set associated with an edge is
	not relevant in the definition of the c-reduction. Therefore,
	$<A,B, C,E,D>$,   $<A,B, C,D,E>$, and $<A,C, B,E,D>$ are also c-reductions
	of $\mathcal{C} $.}
	\label{fig:poset-graph}
\end{figure}

\subsection{Preliminary properties}

\vspace{.2in}

Corollary~\ref{cor:initial-2-c} is used to prove the following Proposition.

\begin{proposition}[Initial species property]
	\label{lem:single-source-more-characters}
	Let $\grbm$ be a skeleton graph and let $T$ be a tree solving $\grbm$. Then
	at least one of the following properties holds:
	\begin{itemize}
		\item
		      the initial state of $T$ consists of a single inactive character $c$,
		\item the tree $T$ is a \simple branch-tree and all inactive
		      characters of the initial species of tree $T$  are lost in distinct
		      branches.
	\end{itemize}
\end{proposition}

The next lemma  shows that sources of active chains of a graph \grbe cannot be internal nodes of any tree $T$ solving \grbe.
This Lemma will be crucial to show that there are at most two active chains in a graph.

\begin{lemma}[sources of chains are leaves or initial states]
	\label{lem:nosource-is-internal-species}
	Let $\grbe$ be an active skeleton graph and let $T$ be a tree solving $\grbe$.
	Let $s$  be a source of  a nontrivial active chain $\cal C$ of the graph
	$\grbm$ induced by $\grbe$.
	Then $s$ is  the initial species or a leaf  of tree $T$ solving the skeleton graph.
\end{lemma}

\subsection{Safe species are sources of active chains: a necessary and sufficient condition}
\label{sec:necessary-condition}

In this section, we show that  a necessary and sufficient condition for a species $s$ to be safe for (active) skeleton graphs is that $s$ is the source of an active chain: this property will be stated in  Corollary
(\Cref{cor:property-necessary-sufficient}).
 
\subsubsection{Safe species are sources of active chains: a necessary condition}

\begin{lemma}
	\label{lem:basic-safe-trees}
	Let $\grbe$ be an active skeleton graph that is  not \degenerate, and let $s$ be a safe   species of $\grbe$.
	Then $s$ is the source of an active chain.
\end{lemma}

Let us recall that a safe species  $s$ must be \pendant in the graph, that is $s$ does not include any other species as it must be realized in the graph.

\subsubsection{Sources of  active complete chains are safe species: a sufficient condition}

For complete active chains  it is easy to show that their sources are safe species, leading to the following sufficient condition.
 
\begin{lemma}
	\label{lem:property-pre-necessary0}
	Let $\grbe$    be an active skeleton graph.
	Let  $s$ be the source of an active complete chain 	$\cal C$ of $\grbe$. Then $s$ is safe for \grbe.
\end{lemma}

\begin{figure}[htb!]
	\centering
	\includegraphics[width=0.2\textwidth]{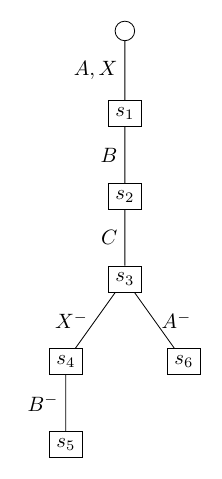}
	\caption{The tree solving the active skeleton graph  of Figure~\ref{fig:poset-graph2}. Observe that the  active chain consisting of  species $<s_1, s_2, s_3>$ is a complete chain and gives the initial-path of the tree.}
	\label{fig:poset-tree}
\end{figure}

Figure~\ref{fig:poset-graph2} illustrates an example of active complete chain with source $s_1$ that is safe. Indeed the tree represented in Figure~\ref{fig:poset-tree} is solving the active skeleton graph of Figure~\ref{fig:poset-graph2}.

\subsubsection{Sources of  active incomplete chains are safe species: a sufficient condition}

In this section we consider the case of a graph having only  incomplete chains and a  main question  related to this case  is  counting the number of  active chains for the diagram of  a skeleton graph, mainly:
\noindent
{\bf how many safe species and active incomplete chains has a skeleton graph?}.

Answering this question we show the following:
\begin{itemize}
	\item if  a skeleton graph is solved by {\bf an \incomplete branch-tree},  then it has {\bf a unique \incomplete  active chain}, i.e. {\bf a unique safe species} (see Lemma \ref{lem:property-branch-incomplete}),

	\item  if  a skeleton graph is solved by {\bf a line-tree}, then  either the graph has  {\bf  two active chains  and the set $A$ of active characters is empty}, or  it has {\bf a unique active chain and $A$ is not empty} (see Lemma \ref{prop:property-line-sufficient}).
\end{itemize}


\begin{lemma}
	\label{lem:property-pre-necessary2}
	Let $\grbe$ be a skeleton graph that has only  incomplete active   chains.
	Let  $s$ be the source of an incomplete active   chain 	$\cal C$ of $\grbe$. Then $s$ is safe for \grbe.

\end{lemma}

\begin{proof}
	Two cases are possible. Case 1): the skeleton graph is solved by an \incomplete branch-tree or 2) is solved by a line-tree. The case 1) is stated in the following lemma~\ref{lem:property-branch-incomplete}, while case 2) is stated in the following lemma
	\ref{prop:property-line-sufficient}.
\end{proof}

\begin{lemma}
	\label{lem:property-branch-incomplete}
	Let $\grbe$ be an incomplete skeleton graph solved by a branch-tree.
	Then the graph has only one active   chain and its source $s$ is the only safe species in the graph.
\end{lemma}

In the following we will use the  notion of \emph{inverted-tree}.

\begin{definition}[inverted-tree]
	Let $T$ be a tree solving a skeleton graph \grbM with initial-species $s_0$
	and $s$ a leaf of tree $T$. Then a skeleton graph \grbM is solved by  an  \emph{inverted-tree}  of tree $T$, if  given tree $T'$ where the path from $s_0$ to $s$ of tree $T$ is replaced by the path obtained by listing all the  species  from $s_0$ to $s$ in inverse order, that is from $s$ to $s_0$,  then $T'$ also solves the skeleton graph \grbM. 
\end{definition}

\begin{figure}[htb!]
	\centering
	\includegraphics[width=0.8\textwidth]{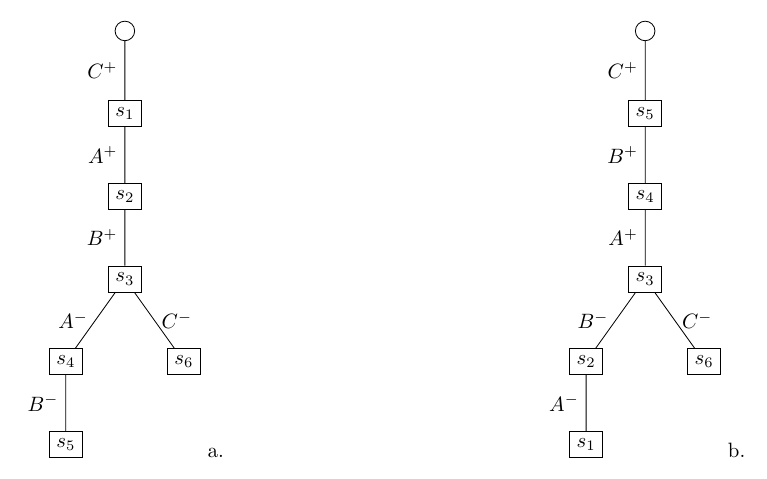}
	\caption{A tree $T$ with a path with species $s_1, s_2, s_3, s_4, s_5$  on the left and the inverted-tree $T'$ of tree $T$ on the right. Both trees have the same set of species and solve the same skeleton graph.}
	\label{fig:inverted-tree}
\end{figure}

\begin{example}
Figure \ref{fig:inverted-tree} illustrates two trees $T$ and $T'$ that solve the same skeleton graph. Observe that a tree $T$ solving a skeleton graph may not be solvable by an inverted-tree.
\end{example}



The next property refers only to  active skeleton graphs with $A$ empty. Indeed, if $A$ is not empty, then the active charaters must be negated along the line-tree  $T$ solving the graph (otherwise the active character is red universal) meaning that the graph cannot be solved by the inverted-tree of the line-tree $T$.

\begin{lemma}
	\label{prop:property-line-sufficient}
	Let $\grbm$ be a skeleton  graph that is solved by a line-tree.
	Then the graph has only two active chains that may be both incomplete  or complete.
	The two active chains have sources that correspond to the initial state of two trees $T$ and $T'$ solving the graph where $T'$ is the inverted-tree of $T$.
\end{lemma}



The following statement is consequence of Lemma~\ref{lem:basic-safe-trees}
(which gives a necessary condition for a species to be safe) and
Lemmas~\ref{lem:property-pre-necessary0} and~\ref{lem:property-pre-necessary2}
which provide respectively a sufficient conditions for complete and incomplete chains.

\begin{corollary}[necessary and sufficient conditions for safe species]
	\label{cor:property-necessary-sufficient}
	Let $\grbe$ be a skeleton graph that is not \degenerate, and let $s$ be a
	species of \grbe.
	Then $s$ is safe for $\grbe$  if and only if $s$ is the source of an active chain of the Hasse diagram for $\grbm$.
\end{corollary}

The following result is a consequence of Proposition \ref{prop:traversal}   stating that  a reduction produces a traversal of the initial path of a tree $T$ solving a graph,   and the fact that the active chains of the graph provides the initial part of reductions as stated in   Lemma~\ref{lem:property-branch-incomplete}  and Lemma \ref{lem:property-pre-necessary0} (if the tree $T$ is a branch-tree) or by Lemma~\ref{prop:property-line-sufficient} (if the tree $T$ is a line-tree).

In other words, any active chain of a graph $\grbm$ is the initial path of a tree solving the graph as stated below.

\begin{corollary}
	\label{cor:active-initial-path}
	The active chains of a graph $\grbm$ consist of the sequence of species of the initial path of a tree $T$ solving the graph $\grbm$.
\end{corollary}

\subsubsection{Safe species in \degenerate graphs}

By definition \ref{def:degenerate} we can show the following fact.

\begin{lemma}\label{lem:all-safe}
Let $\grbe$ be a skeleton graph that is  \degenerate. Then any species of the graph that can be realized is  safe.
\end{lemma}
\begin{proof}
By definition, the realization of a species in the graph leads to the removal of the species from the graph. By definition of \degenerate it is immediate to verify that after the realization of a species the red-black graph has no active characters.
Thus by Proposition \ref{prop:subset-reducible} the graph is reducible, proving what required.
\end{proof}

\subsection{Computing safe species in polynomial time}

The following proposition is used to show the main Theorem \label{teo:polynomial} stating that computing safe species of an active  skeleton graph can be done in polynomial time.

We need first to state the following:

\begin{lemma}
\label{lem:testing-realization}
Testing whether the  realization of a chain in a graph does not induce red $\Sigma$-graphs can be done in polynomial time in the size of the graph.
\end{lemma}
\begin{proof}
By definition the source $s$ of a chain is a \pendant  species which means that any arbitrary order of realization of characters does not remove any species $s'$ smaller than $s$, as $s$ does not  include any species. 
Consequently any order of realization of characters of $s$ produces the same graph. Let $G_i$ be the graph after the realization of the $i$ species of the chain. Now, if the realization of some characters in $C(s_{i+1}) - C(s_i)$ remove a species $s'$ that is included in $s_{i+1}$, it must be that the chain will have $s'$ between $s_i$  and $ s_{i+1}$, a contradiction. Thus any arbitrary order of realization of characters produces the same graph $G_{i+1}$. In other words, to test whether the  realization of a chain in a graph does not induce red $\Sigma$-graphs, it is enough to   choose an aribitrary order of realization of the characters of each  species $s_i$ of the chain, starting from the source of the chain.
The cost of realization of a character is clearly polynomial in the size of the graph and similarly testing each time if the graph induces a red $\Sigma$-graphs can be done in polynomial time.
\end{proof}

Let us now show that also the computation of active chains can be done in polynomial time.

\begin{proposition}
	\label{prop:active-chain}
	Let $\grbe$ be an active skeleton and \grbm its skeleton graph and let $\Hd$ be the Hasse diagram of \grbm.
	Then \Hd has a complete active chain, or all incomplete active chains of \Hd have internal nodes with outdegree $1$. Moreover, an active chain can be computed in polynomial time.
\end{proposition}

\begin{proof}
Clearly, we first need to apply Lemma \ref{lem:testing-realization} to show that testing whether a chain  is active, i.e it can be realized in a graph without inducing red $\Sigma$-graphs,  can be done in polynomial time.
Let us first consider the case that the active chains are complete.  Now, the number of such chains is at most the number of branches of the branch-tree, hence polynomial in the number of characters. By a depth-first visit of the diagram starting from the sink that contain all characters, we can visit the diagram to find the active complete chains.
 
	It suffices to consider the case when all active chains of \Hd  are  incomplete,
	and we need to prove that all their internal nodes have outdegree $1$.
    Let us remind that  by Corollary \ref{cor:active-initial-path}   an incomplete active chain $\cal C$   corresponds to an initial portion of the initial path
	of a tree $T$ solving \grbm.
	Observe that  $\cal C$  consists of species obtained by adding
	characters one after the other. This requires that no
		character is lost in the initial path corresponding to the chain.
	By \Cref{prop:basic-1}, other nodes of the poset are obtained  by negating the first character in the initial path. Thus, it is immediate to show that all other nodes of the poset not in $\cal C$ cannot have incident edges from the chain since the first character of the chain  is not in such nodes thus implying that no internal node of $\cal C$ has outdegree greater than one.

 By the above characterization we can compute the incomplete chains by a depth-first visit of the poset.
	Let $m$ be the number of characters and $n$ the number of species in the graph.
	Now, visiting the graph by a depth-first-visit allows to test candidate active chains in  time which is $O((n + m ) g(m)$, where $O(n+m)$ is the time to visit nodes of the graph while $O(g(m))$ is the cost of realizing a single species node of the chain. Indeed, $O(g(m))$  is a function of the number of characters of a species and represents the time cost of their realization.  Such a cost is at most $O(m \cdot n)$, since it consists of deleting edges outgoing from a character node and eventually adding red edges; and each node species of the chain has at most $m$ characters.
 \end{proof}


Applying the above results we can show the following main Theorem.

\begin{theorem}
	\label{lem:polynomial}
	Let $\grbe$ be an active skeleton graph and \grbm its skeleton graph that is not degenerate.
	Then any safe species $s$ of  graph $\grbe$     can be computed in polynomial time.
\end{theorem}

\begin{proof}
	By \Cref{cor:property-necessary-sufficient} a safe species must be the source of an active chain.
	Thus a safe species is computed by considering active chains of the Hasse diagram for the graph.
	First observe that the construction of the Hasse diagram can be done in polynomial time on the size of the graph as it is based on the comparison between species.    Active chains relie on the characterization given in Lemma~\ref{prop:active-chain}  stating that such chains can be computed in polynomial time. This observation concludes the proof.
\end{proof}

We conclude the section by describing  the main recursive steps for solving (active) skeleton graphs in Algorithm~\ref{alg:Reduction-maximal} {\bf Reduce-Maximal}. The correctness of the recursive Algorithm {\bf Reduce-Maximal} is a consequence of \Cref{lem:polynomial} for non \degenerate graphs or is a consequence of the definition~\ref{def:degenerate} for \degenerate graphs. The polynomial time complexity follows from Theorem~\ref{lem:polynomial}.

\begin{algorithm}
	\caption{Recursive {\bf Reduce-Maximal}. Observe that
			{\bf Realize$(\grbe, S_c)$} is a function that receives in input a graph $\grbe$  and a sequence $S_c$ of charactes and returns the new red-black graph after the realization of the sequence $S_c$ of
		characters}
	\label{alg:Reduction-maximal}
	\begin{algorithmic}[1]
		\REQUIRE{An active skeleton $\grbe$ and let $\grbm$ be the skeleton graph}
		\ENSURE{A  reduction of the graph $\grbe$, if it exists,	otherwise, it aborts.}
		\STATE{${\cal P} $ the Hasse diagram for $\grbm$ }
		\IF{$\grbm$ is not degenerate}
        \STATE $s \gets $ the source of  an active chain of ${\cal P}$
        \STATE $S_c \gets$ the sequence of positive characters of $s$ that are inactive in \grbm 
        \IF{no $s$  exists}
		\STATE{Abort}
        \ENDIF
        \ENDIF
		\IF{$\grbm$ is  degenerate}
        \STATE $S_c \gets$ the sequence of all positive characters  in \grbm
		\ENDIF
		\STATE{	$\grbm \gets$ Realize$(\grbe, S_c)$}
		\RETURN $<S_c, {\bf Reduce-Maximal} (\grbe)>$
	\end{algorithmic}
\end{algorithm}

\section{On the number of trees solving  skeleton graphs}

In this section we show that given non degenerate skeleton graphs then they admit at most two trees solving the graph.
The result is a consequence of  Proposition \ref{prop:at-most-one} and  Lemma \ref{lem:inverted-path} and the characterization in Remark \ref{re:scanonical} of a special branch-tree consisting of only three characters.
In particular, the solution is unique in case it consists of an incomplete branch-tree. In the following we give a characterization of trees solving non degenerate skeleton graphs.

\subsection{Characterization of line-trees}

\Cref{prop:property-line-sufficient} shows that, if a line-tree $T$ solving an
active skeleton \grbe has at least an active character, then \grbe has exactly
one safe   species, therefore $T$ is the only solution.
If \grbe has no active characters, the graph \grbe has at most two active
species $s_{1}$ and $s_2$ that have exactly one (inactive) character.
Those two species $s_1$ and $s_2$ are respectively the initial and final species of a line-tree $T$.
Thus the graph is solved by two trees, $T$ and the tree $T^{r}$  obtained
inverting $T$.

\subsection{Characterization of branch-trees}

In this section we consider the number of possible trees solving skeletons.

The solution is unique in the case the graph $\grbM$ admits an \incomplete branch-tree as stated in Proposition  \ref{prop:at-most-one}. Observe that \Cref{prop:at-most-one} is  a consequence of
Lemma~\ref{lem:property-branch-incomplete} stating that an \incomplete
branch-tree has a unique active chain; clearly it means that there exists a
unique tree $T$ solving the graph.

\begin{proposition}
	\label{prop:at-most-one}
	Let \grbM be an incomplete skeleton graph that is solved by a branch-tree
	$T$.
	Then \grbm has only one safe source.
\end{proposition}

If a skeleton $\grbm$ is solved by a complete branch-tree and it is not
\degenerate, then the Hasse diagram of \grbm  has at most two active chains with
distinct sources, therefore there may be at most two solutions of \grbm with
different safe species: this is stated and proved in Lemma \ref{lem:inverted-path}.


The above Proposition will be proved by the following Lemma~\ref{lem:inverted-path} that details the construction of  the two distinct  solutions of a skeleton graph solved by a non \degenerate  and complete branch-tree having more than three characters. In the case of complete branch-trees having three characters, then  the tree is \ref{re:scanonical} or it is a pseudo-line tree.

\begin{lemma}
	\label{lem:two-complete-active-chains}
	Let $\grbm$ be a complete skeleton graph that is not \degenerate. Let  ${\mathcal C_1}$ and ${\mathcal C_2}$ be two active chains of \grbm
	whose safe sources are respectively $s_1$, $s_2$ and both have more than one character.
Then $s_2$ consists of all characters of
		      $s_1$ except for  one character $a$ of $s_1$ that is replaced by
		      character $b$ in $s_2$, which is a last   gained character in chain ${\mathcal C_1}$.

\end{lemma}
\begin{proof}
	Since \grbm is complete, all its chains are complete, by \Cref{lem:property-branch-complete}. Each of such chain  is the initial path of a tree solving the graph by Corollary \ref{cor:active-initial-path}.
	First assume that $s_1$ is the initial species of a tree $T_1$ solving the graph $\grbm$ and having initial path the chain  ${\mathcal C_1}$.
    Each character of the source $s_1$, by the main
    Proposition~\ref{prop:basic-1}  is lost in a
    distinct branch, therefore any other species (included $s_2$) that is not in the chain ${\mathcal C_1}$ must have
    all characters of $s_1$ except for one --- let us call $a$ such a character.
    Indeed, no two characters of $s_1$ can be both negated along the same branch.
	Symmetrically, $s_1$ has all characters of $s_2$ except for one  --- let us call $b$ such a character.
	Since $s_1$ and $s_2$ are both sources of the Hasse diagram, neither of them
	can be included in the other one and they cannot include  other species.
	This implies that  $s_{2}$ includes all characters of $s_1$ except for a character $a$ that
	is replaced by $b$.
 Let us first  show that $C(s_2 ) = C(s_1) \setminus \{a\} \cup \{b\}$ where $b$  is one of the character added along the chain ${\mathcal C_1}$.
 If on the countrary $  C(s_1) \setminus \{a\} \cup \{b\}\subset  C(s_2 )$, it must be that there exists a character $c$ in $C(s_2) \setminus C(s_1)$ that is distinct from $b$. Now, $c$ and $b$ are gained below the source $s_1$ along the chain ${\mathcal C_1}$ of the tree $T_1$. Both characters are maximal ones and they are not comparable but they are not disjoint as they share species $s_2$.
 They are a confliting pair in $s_2$ as the species $s_1$ induces the configuration $(0,0)$ for the pair and the other configurations are due to the fact they are not disjoint but are not comparable. It follows that the realization of $s_2$ in the graph induces the red $\Sigma$-graph by Lemma \ref{lem:conflict-pair} which contradicts the definition of $s_2$ as  a safe species of the graph.
It follows that $s_2$ consists of all characters of
		      $s_1$ except for  one character $a$ of $s_1$ that is replaced by
		      character $b$ in $s_2$. 
 To conclude the proof of the Lemma, let us show that character $b$ must be the last one added to the chain ${\mathcal C_1}$. Indeed, let $d$ be a character that is the last one added to the chain ${\mathcal C_1}$ and is not in $C(s_2)$. Consider now $(d, b)$. This pair is conflicting in the species $s$ that precedes $s_2 $ in the branch of tree $T_1$. Indeed, the species $s_1$ is a $(0,0)$ configuration while there are two configurations: $(0,1)$  is given in any species $s'$ before the introduction of $d$ in the initial path of tree $T_1$ and the other with configuration $(1,0)$ is the $s_1$ species that introduces $b$ above $d$ in the initial path of tree $T_1$ corresponding to chain ${\mathcal C_1}.$ Now, after the realization of $s_2$, the pair  $(d, b)$ is still conflicting in $s$. This imlies that the realization of ${\mathcal C_2}$ that has species $s$ leads to a red-$\Sigma$-graph by Lemma \ref{lem:conflict-pair} which contradicts the definition of  chain as being active ${\mathcal C_2}$.
\end{proof}

\begin{lemma}
	\label{lem:two-chains}
	Let $\grbm$ be a complete skeleton graph that is not \degenerate.
	Let  ${\mathcal C_1}$ and ${\mathcal C_2}$ be two active chains of \grbm
	whose safe sources are respectively $s_1$, $s_2$ and both have more than one character.
	The following two statements hold:
	\begin{enumerate}
		\item given consecutive internal species $s_i$ and $s_{i+1}$ of chains ${\mathcal C_1}$ and ${\mathcal C_2}$, they  differ by a single character;
		\item the graph has at most two active chains ${\mathcal C_1}$ and ${\mathcal C_2}$.
	\end{enumerate}
\end{lemma}
\begin{proof}
First observe that by Lemma \ref{lem:nosource-is-internal-species} a safe source is either an initial state or a leaf of a tree.

Let us now prove statement 1) of Lemma,   that is given the active chain ${\mathcal C}_1$ of \grbm,  two consecutive species in the chain differ of a single character.  
 As above assume that $s_1$ is the initial species of a tree $T_1$ solving the graph $\grbm$ and having initial path the chain  ${\mathcal C_1}$.  Similarly, given the other active chain ${\mathcal C}_2$ of \grbm, it has source $s_2$.
By Lemma \ref{lem:nosource-is-internal-species} we have that $s_2$ is a leaf of tree $T_1$ and we will use this fact below.

	Assume to the contrary that the chain ${\mathcal C}_1$ has two adjacent internal species $s'$  and $s''$ in the chain such that they differ at least of the two characters
	$d_1$ and $d_2$.
Now, the source $s_1$ is a species that induces a $(0,0)$ configuration for the pair $(d_1, d_2)$ of characters, as  $d_1$ and $d_2$ ara gained in $s''$.
 Observe that by Proposition~\ref{prop:basic-1}   $d_1$ and  $d_2$ are lost on distinct branches of the umbrella of tree $T_1$ which implies that the source $s_2$ (it is a species that is not in chain ${\mathcal C}_1$ but is in the umbrella of tree $T_1$) must necessarily have one of $d_1$ or $d_2$. 
 More precisely, two cases are possible, case 1)   $s_2$ has only  the character $d_1$, case 2) $s_2$ has both characters $d_1$ and $d_2$, since they are lost on other branches.

 Since $d_1$ is not in $s_1$ and is in $s_2$ and by previous Lemma \ref{lem:two-complete-active-chains}   $s_2$ differs from $s_1$ by  a single character, it follows that such character is exactly $d_1$, which is a contradiction with the previous assumption that $s_1$ is a species that induces a $(0,0)$ configuration for the pair $(d_1, d_2)$ of characters. Thus case i) is not possible.

 Thus let consider case 2), that is the pair $(d_1, d_2)$ of characters is in $s_2$. Again we obtain a contradiction since when $s_2$ is realized in the graph the pair is a conflicting pair in the graph as there are species that induce the all four configurations. Again we obtain a contradiction and case 2) is not possible. Since both cases are not possible we have proved statement 1) of the Lemma.

 Let us now prove statement 2) of the Lemma. By applying the previous Lemma \ref{lem:two-complete-active-chains} it must be that $s_2$ is obtained by adding one of the last gained characters to $s_1$ in place of a character of $s_1$.
 This fact implies that all characters along the chain ${\mathcal C_1}$ are lost in a branch of the tree $T_1$ having leaf the safe species $s_2.$  Indeed, let us recall that $s_2$ is assumed to be the leaf of tree $T_1$ and $s_1$ the initial state of tree $T_1$.
 Clearly, by construction there cannot be another leaf species obtained by deleting internal characters gained in chain ${\mathcal C_1}$ which is the initial path of tree $T_1.$ This fact proves that there exists at most two safe species and thus two active chains.
 \end{proof}
 
\begin{figure}[bt]
	\centering
	\begin{subfigure}[b]{0.45\textwidth}
		\includegraphics[scale=0.6]{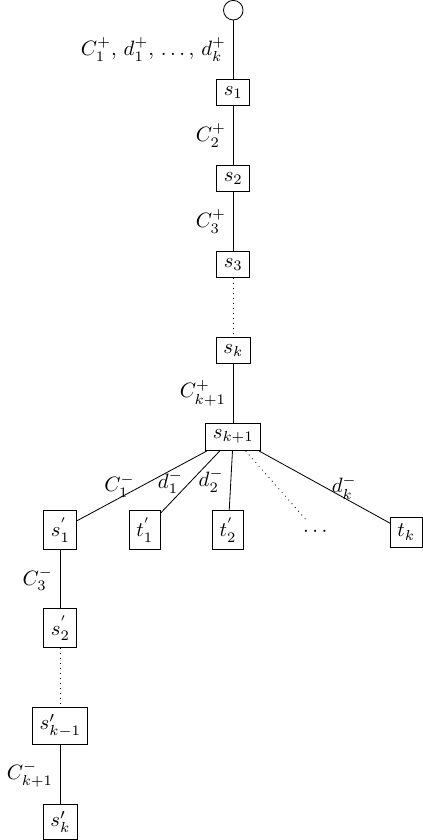}
		\caption{Solution 1}
		\label{sol1}
	\end{subfigure}
	\hfill
	\begin{subfigure}[b]{0.45\textwidth}
		\includegraphics[scale=0.6]{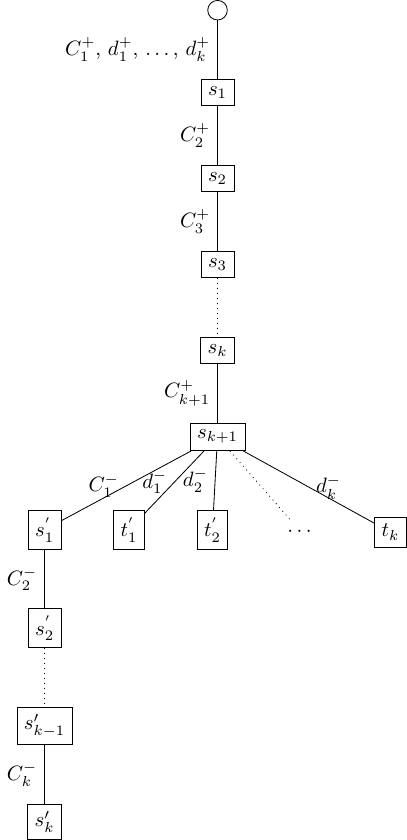}
		\caption{Solution 2}
		\label{sol2}
	\end{subfigure}
	\caption{The figure illustrates the proof of statement 2) of Lemma~\ref{lem:two-chains}. Observe that the solution 1 on the left leads to the leaf species $s'_k$ with characters
 $\{ C_2, X\}$, for $X=\{d_1, \cdots, d_k\}$. Species $s'_{k-1}$ induces the conflicting pair $( C_2, C_{k+1})$ in the graph obtained after the realization of $s'_k$. Only  leaf species of solution 2, i.e. $s'_k$ with characters $\{ C_{k+1}, X\}$ is a safe species.}
	\label{fig:simple-branch-multiple}
\end{figure}

See Figure \ref{fig:simple-branch-multiple} illustrating the Lemma  \ref{lem:two-chains}.

As a consequence of the previous Lemma the following fact holds.
Observe that  the statement of the Lemma below excludes special branch-trees that solve skeleton graphs consisting of three characters. Indeed, the Lemma shows that if the graph has more than three characters and is complete, then it has at most two safe species, each having at least two characters. As a consequence, the previous Lemma \ref{lem:two-chains} applies.
Such graphs are called \scanonical  and are illustrated in the next example.

\begin{example}
    \label{re:scanonical}
    A special type of skeleton graph is the one having
    exactly three characters \texttt{A}, \texttt{B}, \texttt{D} and 6
    species whose characters are $\{\mathtt{B}\}$, $\{\mathtt{D}\}$,
    $\{\mathtt{A}, \mathtt{B}\}$, $\{\mathtt{A}, \mathtt{D}\}$,
    $\{\mathtt{B}, \mathtt{D}\}$, and
    $\{\mathtt{A}, \mathtt{B}, \mathtt{D}\}$.
    In this case the solutions are two branch-trees (see
    Figure~\ref{fig:canonical-forest1}) that are obtained one from the other
    by swapping the two character \texttt{B} and \texttt{D}.
    Observe that the graph   has at most two  safe  species $s_1$ and $s_2$ with singletons.
\end{example}

\begin{lemma}
	\label{lem:inverted-path}
	Let $\grbm$ be a complete skeleton graph  that is not \degenerate and has
	more than three characters and is solved by a branch-tree.
	Then $\grbm$ has at most two  safe  species $s_1$ and $s_2$ that are the
	initial species of two solutions (respectively $T_1$ and $T_2$) of \grbM.
	If such two species $s_1$ and $s_2$ exists, then
	(1) $C(s_1)$  differs from $C(s_2)$ by a single character $a$ which is replaced by  $b$ in $C(s_2)$ (2)
	$s_2$ is a leaf of the tree  $T_1$ and $T_2$ is obtained by inverting the path of $T_1$ from $s_1$ to $s_2$.
\end{lemma}
\begin{proof}
	Assume first that $s_1$ consists of a single character $c$ and let $T_1$ be the tree with initial species $s_1$. Then we show that we obtain a contradiction and thus Lemma \ref{lem:two-chains} must apply.  Since the graph is complete, the chain with source $s_1$ will consists of species obtained by the gain one after the other of each character of the graph.  Observe that since $s_2 \not= s_1$, it must be that $s_2$ will not have character $c$, otherwise the chain with source $s_2$ will start with $s_1$.  On the other hand $s_2$ will not have the last gained character $d$ of the chain with source $s_1$, otherwise the tree solving the graph whose initial path is the chain with source $s_1$ must be a  pseudo-line tree with $d$ negated on a single branch with such negation  (let us recall that characters are negated in the order they are introduced in the three by Proposition \ref{prop:basic-1}).  It follows that $s_2$ consists only of  internal characters of the the tree $T_1$, which implies that $s_2$ consists of a singleton, as any pair of internal characters are a conflicting pair (they are not comparable and $s_1$ is the $(0,0)$ configuration for such pair).
	Observe that given species $s'$ consecutive to $s_2$ in the chain $C(s_2)$,  it is either the same species consecutive to $s_1$ in chain $C(s_1)$, which implies that the tree has three characters at
    most (see Figure \ref{fig:canonical-forest1} and example \ref{re:scanonical}) illustrating this case), or otherwise it is a species having a charcater $e$ and it is
    immediate to show that $(c,e)$ is a conflicting pair, which is a
    contradiction. Thus $s_1$ consists of more than one character. Then by can apply Lemma \Cref{lem:two-chains}  and Lemma \ref{lem:two-complete-active-chains} and consequently  a reducible graph $\grbm$ cannot
	have a third active chain if it is not \degenerate.
	Given ${\cal C}$  and ${\cal C}'$ the two chains of the graph, observe that if characters of the initial path of tree $T$ form chain ${\cal C}$, then ${\cal C}'$  is obtained by the negation of characters gained in the initial path except for the last one, proving the Lemma.
\end{proof}

\subsection{Characterization of type of trees solving skeleton graphs}
\label{sec:characterization}

Based on the results of the previous section the following cases are possible, as  stated by \Cref{thm:skeleton with two safe species}.

\begin{figure}[tb!]
    \includegraphics[width=.6\linewidth]{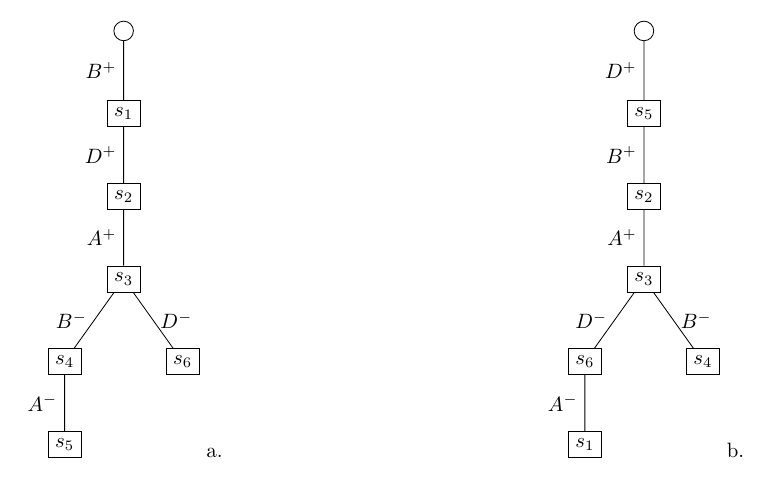}
    \caption{The \scanonical skeleton and its two solutions. Observe that such
        two trees are isomorphic, since they can be obtained by swapping  the
        two character \texttt{B} and \texttt{D}}
    \label{fig:canonical-forest1}
\end{figure}

\begin{theorem}
    \label{thm:skeleton with two safe species}
    Let \grbm be a skeleton graph that has two  or more safe species.
    Then at least one of the following conditions holds:
    \begin{description}
        \item[\scanonical]
            \grbm has exactly three characters \texttt{A}, \texttt{B}, \texttt{D} and 6
            species whose characters are $\{\mathtt{B}\}$, $\{\mathtt{D}\}$,
            $\{\mathtt{A}, \mathtt{B}\}$, $\{\mathtt{A}, \mathtt{D}\}$,
            $\{\mathtt{B}, \mathtt{D}\}$, and
            $\{\mathtt{A}, \mathtt{B}, \mathtt{D}\}$.
            In this case the solutions are two branch-trees (see
            Figure~\ref{fig:canonical-forest1}) that are obtained one from the other
            by swapping the two character \texttt{B} and \texttt{C} (see Figure \ref{fig:canonical-forest1} for trees solving such graph);
        \item[\degenerate]

            \grbm is \degenerate with $m$ characters  $c_1, \dots , c_m $, with $m \geq 3$.
            In this case, the graph is solved by all branch-trees $T_{i}$ (with
            $1\leq i\leq m$) consisting of an initial state with the set
            $D=\{c_{j}, i \neq j\}$, and an umbrella whose branches correspond to the
            loss of a character of $D$ (see \Cref{prop:degenerate});
        \item[2-solvable]
            \grbm is solved by two branch-trees that differ only for the inversion
            of a path from the root to a leaf, called {\em inverted-path} --- the safe
            active  species are called {\bf sources} of the graph (By
            \Cref{lem:inverted-path});
        \item[line-2-solvable]
            \grbm it is solved by two line-trees that differ  only for the inversion of
            a path from the root to a leaf, called {\em inverted-path} --- the safe
             species are called {\bf sources} of the graph;
    \end{description}
\end{theorem}
\begin{proof}
Let us recall that a skeleton graph is solved either by a branch-tree or a line-tree based on Lemma \ref{lemma:skeleton-shape}.
Observe that each of the following conditions are mutually exclusive.  Indeed, either the graph is \degenerate or is not \degenerate. If it is \degenerate then  Proposition \Cref{prop:degenerate} applies. Otherwise,  if it is not \degenerate, either the graph is solved by a branch-tree or by a line-tree. In the first case, if the branch-tree has more than three characters then Lemma \Cref{lem:inverted-path} applies (2-solvable). Otherwise if it has three characters, it may be \scanonical or it is a unique solution. Otherwise, if the graph is solved by a line-tree then  it is solved by at most two lines trees (line-2-solvable).

\end{proof}

The following is a consequence of \Cref{prop:at-most-one} and \Cref{thm:skeleton with two safe species}.

\begin{corollary}
	\label{cor:lisskeleton tree-trees}
	Let \grbm be a skeleton graph.
	Then one of the following two cases holds:
	\begin{enumerate}
		\item
		      \grbm has at most one solution and its initial character is obtained
		      from its initial species;
		\item
		      \grbm satisfies one of the cases listed in
		      \Cref{thm:skeleton with two safe species}.
	\end{enumerate}
\end{corollary}

By abuse of the language, the skeleton tree $T$ solving a 2-solvable skeleton graph \grbm is also called 2-solvable meaning that the tree $T'$ obtained from $T$ by posing the leaf source of $T$ as root of the new tree $T'$ is also another skeleton tree solving \grbm.


Observe that a main property of a reduction for a graph is the following:

{\bf Property reduction:}
Given a reduction, any subsequence consisting of $X$  characters is a reduction for the subgraph induced by the set $X$ of characters.

\section{PART II:  solving general graphs}

The main observation that leads to a polynomial time algorithm for solving general graphs is the following: after the realization of the characters of the source of the skeleton graph of a general graph, if the graph has active characters, then there exists a unique   safe species of  the active skeleton graph, thus the choice of characters to be realized is unique.
This fact implies that there is only one way to reduce the graph either 1)  to a reducible graph without active characters or 2) to a graph with the forbidden configuration represented by the red $\Sigma$-graph.
Assuming that the initial graph is reducible, then the graph we obtain without active characters is still reducible and again we continue the iteration of realizing a safe species of the active skeleton graph. 
Otherwise if we obtain a forbidden configuration, it means that we did a wrong choice of a safe species in the active skeleton and thus we can try to reduced the graph with the other safe species. 
Observe that these iterations correspond to explore a   tree consisting of a single $m$ long path (being $m$ the number of characters) where some nodes may have degree at most $m$ and are roots of subtrees consisting of   at most $(m-1)$-long paths  corresponding to exploration of reductions starting from  $m$ possible safe species of the active skeleton. The single $m$ long path corresponds to the correct reduction of the tree, while the other substrees correspond to erroneus reductions from  the wrong safe species. Each substree has at most $m^2$ nodes and there at most $m$ of such subtrees. Thus the exploration of the tree is polynomial in the size of the graph. 

Clearly, if the initial graph is not reducible then the exploration of the tree leads only to graphs with the forbidden configuration rapresented by the red $\Sigma$-graph or to a graph with a skeleton without safe sources.

Let us state the main Proposition that leads to the polynomial time algorithm to solve general graphs.
Recall that by Theorem \ref{lem:polynomial} we can compute in polynomial time safe sources for the   skeleton graph, respectively active skeleton graph.

\begin{proposition}
	\label{pr:unique-active}
	Let \grb be a reducible graph and $s$ a safe source for the active skeleton graph $\grbe$ of \grb.
	Then given the active skeleton graph $G'$ of the graph   obtained after the realization of the safe source $s$ in \grb,  then $G'$ has a unique safe source.	
	\end{proposition}

\begin{proof}
	Given $s$  a safe source for the active skeleton graph $\grbe$ of \grb, let $G_{RB}(s)$ be the graph obtained after the realization of the   species $s$  in \grb. First, let us show that the active skeleton graph must have active characters. Indeed, if it does not have active characters it means that the characters of $s$ are disjoint from those in teh graph after the realization of $s$ which is not possible since characters are maximal ones and the graph is connected by definition.
    If   characters of  $s$ are active in $G_{RB}(s)$, it follows that they will need to be negated in $G_{RB}(s)$ after some realizations of characters.  This fact implies that all the other species that were safe sources in $\grbe$ (note that they may not be safe in \grb, instead) will have red incoming edges in $G_{RB}(s)$ thus implying that are not feasible anymore in in $\grbe$, i.e. they cannot be removed from in $\grbe$, as shown below.
More precisely, let us distinghish the different types of solutions stated in Theorem \ref{thm:skeleton with two safe species}. Case 1: assume the graph \grb has a skeleton that is 3-canonical. Clearly, the skeleton graph will have a unique safe species that is the one below the root of the tree.  Case 2: assume the graph \grb has a skeleton that is m-canonical. Clearly, the skeleton graph will have a unique safe species that is the species with all characters. All the other species of the graph indeed, will have incoming red edges.
Case 3: assume the graph \grb has a skeleton that is 2-solvable. Since the trees solving the skeleton graph  differ for the inverted-path,  Lemma \ref{lem:inverted-path}, it follows that the leaf of the tree having root the safe species $s$ that has been realized is not feasible (it has red incoming edges) and thus the safe species is unique in this case.
Case 4: assume the graph \grb has a skeleton that is line-2-solvable. Since the trees solving the skeleton graph  differ for the inverted-path, it follows that the leaf of the tree $T$ having root the safe species $s$ that has been realized is not feasible (it has red incoming edges) and thus the safe species is also unique in this case.
All the above cases prove the Proposition.
	\end{proof}

\begin{algorithm}
\caption{Reduce-general$(G,A)$}
	\label{alg:PPR}
\begin{algorithmic}[1]
\REQUIRE{A red-black graph $G$ with set $A$ of active characters}
		\ENSURE{An empty graph,  if $G$ is reducible, otherwise FAIL}
\STATE status : = not FAIL\;
\IF{Graph $G$ is empty}
\STATE return successful\;
\ENDIF
		\IF{Graph $G$ is not empty and status not FAIL}
		\STATE Compute the  skeleton $\grbM$\;
		\STATE    $L_r$ :=  list of safe species of $\grbM$\;
\FORALL{ $r \in L_r$}
\STATE status := not FAIL \;
\STATE realize $r$  in $G$\;
\STATE remove $r$ from $L_r$\;
		\STATE   update $G$ and the set $A$ of active characters\;
\FORALL{$G_i$ connected component of $G$}
		\STATE    status := Reduce-general$(G_i,A)$\;
		\ENDFOR
        \ENDFOR
  \ENDIF
  \IF{$L_r$  is   empty and $G$ is not empty}
  \STATE return FAIL\;
 \ENDIF
	 	
\end{algorithmic}
\end{algorithm}


Let us  give an outline of the main steps of the  polynomial time  algorithm for solving reducible general red-black graphs. The correctness and computation time are proved in Theorem \ref{th:correctness-general}.

Given a red-black graph \grb  the skeleton graph \grbm is computed.  If \grbm is \degenerate then all maximal characters of the graph are realized in any order. 
Otherwise  we need to compute a safe species $s$ for the active skeleton graph. Such a safe species may not be safe for the general graph. Clearly, if $s$ is unique there no  choice, otherwise   based on the application of  Proposition \ref{pr:unique-active} there is a unique safe species in the active skeleton after a safe species is realized in the graph. Correctness of the Algorithm \ref{alg:PPR}, as stated before,   is proved in the next Theorem.

\begin{theorem}
    \label{th:correctness-general}
    Algorithm \ref{alg:PPR} computes a reduction of a reducible graph in polynomial time.
\end{theorem}

\begin{proof}
The main algorithm \ref{alg:PPR} iterates the realization of a safe species in $L_r$ of the active skeleton of the graph $G$ that has to be reduced.

Such iteration is given by line 8 of the algorithm \ref{alg:PPR}.
  The correctness of this step for $L_r$ not empty,  is given by the invariant property stated in Proposition~\ref{prop:maximal-first}, by which there exists a reduction $R$ of $\grb$ that starts with a maximal inactive character $c$ of the skeleton graph.
   
 Now,  we focus on finding the safe species,
\ref{def:safe}) of the Hasse diagram that determines the first characters of skeleton graphs to
realize. Indeed, by Proposition \ref{prop:initial-species-extended} any tree solving an active skeleton graph must have a safe species.  Moreover, in virtue of Lemma \ref{lem:all-safe} we do not distinguish the case of \degenerate graphs.
 Observe that the computation of a safe species is polynomial in the size of the graph based on the statement of Proposition \ref{prop:initial-species-extended}.
Observe that algorithm \ref{alg:PPR} has a main iteration from lines 8-14. In particular line 9 consists of considering a species $r$ in the list $L_r$ of safe species. Then $r$ is realized in graph $G$ which is updated and the procedure is recursively called for the  connected components $G_i$ of graph $G$.
By the main Proposition \ref{pr:unique-active}, after chosing $r$ and realizing $r$ in graph $G$ (line 9), in the next steps there always exists a unique   safe species in the skeleton of the updated graph $G$. If the realization of $r$ leads to a red $\Sigma$-graph in the updated graph and $L_r$ is not empty, it means that the original
 selection of $r$ was wrong, we should fail in reducing a graph (see Line 17-18) in which case we restart the iteration from line 8 and then try the selection of a new safe species in $L_r$ and repeat the realization from the original graph $G_I$.
 Clearly, we have two conditions  1)  either the graph $G$ will be reduced to the  empty one, or 2) $G$ is not empty but $L_r$ is empty. Condition 2) leads to  exit the procedure with FAIL, meaning that all the safe sources have been tried, or no safe species is given in the updated graph.

Clearly, the number of times we repeat  the reduction starting from the same graph but a different safe species is related to the size of $L_r$ for a given connected component.
 By the main characterization stated in Theorem \ref{thm:skeleton with two safe species}, in the worst case  we have to try   $m$ candidate safe sources and for each such sources  the graph may be updated after the realization of  at most $m$ characters. Thus  in the worst case the cost of all iterations is at  most $O(m \times m \times f(n))$, where $m \dot f(n)$ is the time cost for realizing $m$ characters in the graph ($f(n)$ the cost of realization of a single character).  More precisely,  all iterations of the Algorithm \ref{alg:PPR} starting at line 8 may be represented by an exploration tree. This tree consists of  a unique $m$ long path  of nodes $i$, for $i \in \{1, \cdots, m\}$ that represents the sequence of realization of a reduction if the graph is reducible and then there are additional paths that represent the alternative exploration of other candidate realizations. Indeed, each node $i$ in this long path represents the for iteration at line 8 for the updated graph $G$ (after the realization of a safe species) with at most $m-i+1$ characters. Then  node $i$   is the root of a tree consisting of at most $L_r(i) $  simple paths, each path corresponding to a sequence of at most  $m- i +1$ characters that has to be realized. Indeed, whenever  we have $m-i+1$ safe sources (see the \degenerate case), as stated in  the characterization of Theorem \ref{thm:skeleton with two safe species}, we may try all safe species to reduce the graph. Thus the $L_r(i) $  simple paths represents the different sequences of realizations to reduce the graph that need to be explored. Only one of such path will lead to reduce the graph to the empty one, and this is exactly the long path.
  Let us recall that by Theorem \ref{thm:skeleton with two safe species} there are at most two safe species to try each time, unless we try $m- i +1$ possible species only once.
  The size of the exploration tree is at most cubic in $m$, as  the two safe species induce a unique reduction and one safe species is the correct one. Since the realization of a character in the graph costs   at most  $O(n)$, we obtain a total time that is $O(m^3 \times n)$.
Thus the total  time is polynomial in the size of the graph, as required.

\end{proof}

To conclude the section, if the general graph is not reducible applying Algorithm \ref{alg:PPR}, the algorithm will lead to a fail status   due to the absence of a safe species for the graph.

\section{Proof of section~\ref{sec:m-characters}}

\vspace{.2in}

 {\bf Proof of Proposition~\ref{prop:maximal-first}}
 {\em Let \grb be a reducible red-black graph and let \grbM be its skeleton graph.
	Then there exists a reduction $R$ of $\grb$ that starts with a maximal inactive character $c$.
	Moreover, there exists  a reduction of \grbM that also starts with $c$.}
\begin{proof}
	Let \grb a reducible  red-black graph and let $T$ be a tree solving \grb in its standard form, i.e. each internal node, except for the root $r$ is labeled by a species or has at least two children.
	Let $x$ be the only child of the root of $T$. By Proposition~\ref{prop:traversal}, the reduction of a graph starts arbitrarily from characters labeling any edge from the root $r$ of the tree.

	Now only positive characters can label the edge $(r,x)$ --- in fact, a
	negative character $d^{-}$ labeling $(r,x)$ implies that no species in \grb has the character $d$.
	Moreover, by Proposition~\ref{prop:traversal}, positive characters labeling the edge $(r,x)$ can be realized in any order, resulting in the same tree. Therefore, if any inactive maximal character labels the edge $(r,x)$, then it is trivial to obtain a reduction that starts with such maximal character.

	Let $c^{+}$ be a positive character $c^{+}$ labeling $(r,x)$ that maximizes $|S(c)|$ among all characters labelling $(r,x)$. We will prove that $c$ is maximal, hence completing the proof of the proposition.
	Assume to the contrary that $c$ is not maximal. Then there exists an inactive character $c_1$ such that $S(c) \subset S(c_1)$.
	By our choice of $c$, the positive character $c_1^+$ must label an edge in the subtree of $T$ rooted at $x$.
	If $x$ is labeled by a species  $s_x$, then $s_x\in S(c)\setminus S(c_1)$, which contradicts our assumption that $S(c) \subset S(c_1)$.
	Hence, $x$ is not labeled by a species.

	By definition of standard form, $x$ has at least two children $y_1$ and $y_2$. Without loss of generality, assume that $c_1$ labels an edge in the subtree rooted at $y_1$.
	Since $S(c) \subset S(c_1)$, there exists a species $s \in S(c_1)\setminus S(c)$ which implies that the edge labeled by $c^{-}$ is further from the root than the edge labeled by $c_1^+$.
	But this fact implies that the  subtree rooted at $y_2$  has a species $s_2\in S(c)\setminus S(c_1)$, contradicting the assumption that $S(c) \subset S(c_1)$.

	Since the above cases lead to a contadiction,  it must be that $c$ is  a character of \grbM. Therefore removing from the reduction $R$ all character that are not maximal, we obtain a reduction of \grbM starting with $c$, concluding the proof.
\end{proof}

{\bf proof of Lemma~\ref{lem:da-fare}}
{\em Let \grb be a reducible graph. Then  \grbe and each connected component of  
	\grbM is a reducible graph.}
\begin{proof}
	The result is a main consequence of the fact that given a a tree  $T$ solving  \grb then the set of species and characters of \grbe and \grbM induces  trees contained in $T$ showing that also \grbe and \grbM admit a  reduction, i.e. they consist of reducible connected graphs. By a tree contained in $T$ we mean a tree obtained from $T$ by contraction of paths that are deleted from $T$ because of the removal  of species and characters.
	Contracting a path from vertex $x$ to a vertex $y$ in the tree, means removing the path and have vertex $x$ coincident with vertex $y$.
\end{proof}

\section{Proofs of section~\ref{sec:skeleton-graphs}}

{\bf Proof of Lemma~\ref{lem:single-source-more-characters}}
{\em Let $\grbm$ be a skeleton graph and let $T$ be a tree solving $\grbm$. Then
	at least one of the following properties holds:
	\begin{itemize}
		\item
		      the initial state of $T$ consists of a single inactive character $c$
		\item the tree $T$ is a \simple branch-tree and all inactive
		      characters of the initial species of tree $T$  are lost in distinct
		      branches.
	\end{itemize}}

\begin{proof}
	Assume that the initial state has two inactive characters $c_1, c_2$.
	If they are negated along the same path they are comparable, so they must be negated along distinct paths.
	Therefore the tree $T$ has a split-node.
	By Proposition~\ref{prop:basic-1}, they must be negated below the split-node.
	Moreover, by maximality of all characters of \grbm, no character can be
	gained below the split-node, otherwise it would be contained in $c_{1}$ or
	in $c_{2}$.
	Therefore all inactive characters are gained above the split-node.
	By Definition~\ref{def:simple-complete}, the tree $T$ is a  \simple
    branch-tree.

\end{proof}

The following Lemma is an easy consequence of the definition of active chain.

\begin{lemma}
    \label{lem:technical-active-nonactive}
    Let \grbe the active skeleton   of a skeleton graph \grbm. Let  $\mathcal C$ be  an active chain of \grbm. Then the realization of $\mathcal C$  in \grbm does not induce red $\Sigma$-graphs.
\end{lemma}

The above technical Lemma \ref{lem:technical-active-nonactive} will allow to simplify the proofs of the following Lemmas as we will assume an empty set of active characters.  

 In order to prove the Lemma let us recall that a chain is defined over the poset from graph \grbm and a main property of a chain is the following.

 \begin{lemma}
 \label{lem:positive-negative-chain}
Let $\mathcal C $ be a chain of a connected skeleton graph \grbm. Then if the chain   has the gain and then the loss of the same character in two adjacent species of the chain, it must be that an additional  character is gained and then not lost in the two species.
 \end{lemma}

 \begin{proof}
 \Cref{lem:positive-negative-chain} is a consequence of the fact that the skeleton graph \grbm by assumption is connected. Indeed, if a character is gained and then lost and no other character is gained and then not lost in the two species, it must be that the graph is not connected, contradicting the assumption.
 \end{proof}

{\bf Proof of  \Cref{lem:nosource-is-internal-species}}
{\em Let $\grbe$ be a skeleton graph and let $T$ be a tree solving the skeleton graph $\grbm$ induced by $\grbe$.
	Let $s$  be a source of  a nontrivial active chain $\cal C$ of the graph
	$\grbe$.
	Then $s$ is  the initial species or a leaf  of tree $T$.}
\begin{proof}

    We prove the Lemma by contradiction assuming to the contrary that $s$ is neither the initial state nor a leaf of tree  $T$. Observe first that $s$ cannot be a node of the umbrella since  every  edge  in the umbrella is labeled by the negation of characters, and thus non leaf species of the umbrella have incoming edges in the Hasse diagram, thus contradicting that $s$ is the source of a chain. For the same reason  $s$ cannot be the split-node, as in this case $s$ will have incoming edges from species below the split-node as below such node only negated characters occur. Consequently,  $s$ must be a vertex of the initial path of the tree $T_1$ having initial species $s_0$.
    The chain $\cal C$ has at least two species, since $\cal C$ is nontrivial.
    Observe that,  since $s$ is a source of the poset and occurs along the
    initial path of $T$,  $s$  is preceded
    along the initial path of tree $T'$ by a species $s_1$ such that $s_1$ is
    connected to $s$  by an edge labeled by at least the negation of  an inactive
    character, otherwise $s$ will have an incoming edge from species $s_1$, thus contradicting the fact that $s$ is a source of the poset.   
    Moreover, since the chain $\cal C$ is not trivial, there exists another
    species $s_2$ such that $(s, s_{2})$ is an arc of $\cal C$, therefore $s_2$
    must have an inactive character $c_2 \notin C(s)$.
    Observe that, by Proposition~\ref{prop:basic-1}, by which characters are
    negated along the initial path in the order they are
    gained,  the first character to be negated is the one of $s_0$, and thus  $s_1$ will have an inactive character $c$ which is distinct from the one in species $s_0$ and will be gained in a species occuring before $s_1$. Indeed, by Lemma \ref{lem:positive-negative-chain},   $s_1$ cannot be adjacent to $s_0$.
    In the following, we show that   $(c, c_2)$ is a conflicting pair of
    characters  in the species $s_2$ in the graph obatained after the realization of $s$, which leads to contradict the fact that the realization of the active chain does not induce red $\Sigma$-graphs. Indeed, by Lemma \ref{lem:conflict-pair} it will follow that the red $\Sigma$-graph will be indiced by the realizatioon of $s$ and then $s_2$.
    
    First observe that they are not comparable characters being maximal inactive characters. Moreover, the initial species $s_0$ does not have
    any of those characters (\ie it induces the $(0,0)$ configuration for the
    pair $(c,c_2)$).  Observe that there are two species in the graph inducing the configuration $(1,0)$ for the pair $(c,c_2)$, more precisely species $s_1$ and species $s$ that both have character $c$ and not $c_2$.  
Observe that the realization of the chain $\mathcal C$ implies the realization of $s$ and  $(c,c_2)$ is still a conflicting pair in the graph where $s$ is realized.
Indeed, species $s_0$ and $s_1$ cannot be realized before $s$ (neither they are included in $s$ by definition of feasible species, as it must be \pendant). This concludes the proof of the Lemma, and since we obtain a contradiction, it must be that $s$ is not an internal node, but it must be either the leaf of the initial state $s_0$, as required. 
\end{proof}

\subsection{Proofs of subsection~\ref{sec:necessary-condition}}

Let us recall that the red-black graphs in the statement of the Lemmas below are reducible by assumption.
\vspace{.2in}

{\bf Proof of Lemma~\ref{lem:basic-safe-trees}}
{\em Let $\grbe$ be an active skeleton graph that is  not \degenerate, and let $s$ be a safe   species of $\grbe$.
	Then $s$ is the source of an active chain of \grbe.}
\begin{proof}
The Lemma is an easy consequence of the 
 definition of safe species $s$, \ref{def:safe} by which the realization of $s$ produces a reducible graph.

Clearly, since $s$ is the initial species of a reduction (i.e. it leaves a reducible graph)  is also  a source of a tree $T_m$ solving the associated skeleton graph \grbm. Given \grbm and $s$
   we can apply \Cref{lem:single-source-more-characters} stating that either i) $s$ consists of a single inactive character or if $s$ has more characters then ii) the tree is a complete  branch-tree. If case i) applies, then all the species below $s$  in the initial path of the tree  $T_m$ that are obtained by adding characters are in inclusion relation one with each other thus forming a chain $\cal C$ of the Hasse diagram having source $s$.   If instead case ii) applies, the species of the initial path of the tree $T_m$ ordered by the inclusion relation form a chain and in this case $s$ is   necessarily the source of this chain, as it is a \pendant species.

 In both cases, the chains obtained in this way are also active chains in \grbm, i.e. they can be realized in \grbm without producing red $\Sigma$-graph.  Indeed,  again in virtue of Proposition \ref{prop:traversal} the traversal of tree $T_m$ leads to a reduction of the graph to the empty one, and thus no red $\Sigma$-graph is produced with the realization of such chains. Observe that in case i) the chain can be an initial portion of the initial path of $T_m$, while in the case ii) the chain corresponds to the initial path of tree $T_m$ which is complete.
 The similar argument applies to tree $T$ solving the graph \grbe and so we can conclude the proof also for $\grbe.$

\end{proof}



{\bf Proof of Lemma~\ref{lem:property-pre-necessary0}}
{\em Let $\grbe$    be an active skeleton graph.
	Let  $s$ be the source of an active complete chain 	$\cal C$ of $\grbe$. Then $s$ is safe for \grbe.}
\begin{proof}
	By definition of active complete chain the realization of the species of the chain results in a
 red-black graph without  no inactive character, i.e. the graph is empty. Thus the active complete chain directly gives a reduction of the graph and hence its initial species is safe.
\end{proof}


\subsection{Technical Lemmas}

The following   properties give a first characterization of \incomplete branch-trees and are used to prove some of the  lemmas stated below.

\begin{lemma}
	\label{lem:property-branch-incomplete>3}
	Let $\grbm$ be a skeleton graph that is solved by an \incomplete branch-tree and is not a pseudo-line tree. Then the graph has more than three inactive characters.
\end{lemma}
\begin{proof}
	Assume on the contrary that the graph $\grbm$ has only three inactive characters.
	Observe that by Lemma~\ref{lem:single-source-more-characters} the initial state must contain a unique character, which we denote by $b$: then $b$ is the first character to be negated along the initial path in virtue of the main property~\ref{prop:basic-1}.
	Since the graph is solved by an incomplete branch-tree, then $b$ cannot be negated after the gain of all three characters, and hence it will be negated after the gain of the second inactive character $d$. This fact implies that the other two characters, $d$ and $c$, will be negated along distinct branches of the umbrella. Therefore, the branch-tree is indeed a pseudo-line tree.

\end{proof}

\begin{lemma}
	\label{lem:property-inverted-order}
	Let $\grbm$ be a skeleton graph that is solved by an incomplete   branch-tree $T$  that is not a pseudo-line tree and such that the initial state and a leaf species are singletons. Then there exist two characters $c$ and $d$ such that (1) $c$  is gained before $d$ in $T$, (2) both $c$ and  $d$ are lost  in the tree, and (3) $c$ and $d$ are not lost in the same path of $T$.
\end{lemma}

\begin{proof}
	If we assume on the contrary that all characters are lost in the same order they are gained, it is immediate to show that since there exists a path from the initial state to the leaf with only singletons, then such path is either a line-tree or a pseudo-line tree,  a contradiction.
\end{proof}

Let us first state a technical Lemma used to prove Lemma~\ref{lem:incomplete-chain-with-leaf}.
\begin{lemma}
	\label{lem:technical-01-maybe}
	Let $\grbm$ be a skeleton graph and let $T$ be a tree solving the graph. Let
	$s$ be a species that is a leaf of tree $T$ that is the source of an active chain of \grbm  and such that the edge that connects $s$ and its parent $s'$ in $T$ is  labeled by the  character ${d}^-$.
	Let $c$ be a character in  $C(s)$  which is not in the  initial species of the tree $T$ and $c$ is gained before $d$ along the initial path of tree $T$. Then the realization of $(s, s')$  induces the red-$\Sigma$ graph.
\end{lemma}

\begin{proof}
Observe that $s'$ contains the pair of inactive characters $c$ and $d$. 
We can show that the pair  $(c, d)$ is a conflicting pair in the graph $\grbm$ after the realization of species $s$, thus implying that if $s'$ is feasible the realization of the pair $(s, s')$ leads to a red-$\Sigma$ graph, as a consequence of Lemma \ref{lem:conflict-pair}.
	By hypothesis, $c$ and $d$ occur below the initial species $s_0$ and then $s_0$ induces the configuration $(0,0)$  for such pair (indeed $c$ is not in $C(s_0)$) even after the realization of $s$ in graph $\grbm$.  Moreover,   the pair $(c, d)$ induces the $(1,0)$ configuration in  two species of $\grbm$, more precisely  species $s$  and the species $s_2$ where $c$ is gained before $d$. Observe that $s_2$  is still present in the graph $\grbm$ after the realization of $s$.  Finally the configuration $(0,1)$ is induced by the pair $(c, d)$ in a species of the umbrella of tree $T$,  that is in a branch where $c$ must be negated (or lost). Such a negation exists since $c$ and $d$ are not comparable characters and moreover since $c$ is in the species $s$ that is a leaf of $T$, the only way to negate $c$ is in the umbrella of tree $T$. This fact concludes the proof, since the realization of $s$ does not remove in graph $\grbm$ the species inducing  the conflicting pair $(c, d)$ in $s'$.
	\qed\end{proof}

\begin{lemma}
	\label{lem:incomplete-chain-with-leaf}
	Let $\grbm$ be a skeleton graph and  let $T$ be an \incomplete branch-tree that is not a pseudo-line tree  and solves    graph $\grbm$. Let $s$ be a leaf species and
	let $c$ be the unique character in   species $s$  which is not in the  initial species of the tree $T$.
	Then $s$ cannot be the source of an active chain.
\end{lemma}

\begin{proof}
	By Lemma~\ref{lem:property-branch-incomplete>3} the branch-tree has more than three characters.
	Let $s$ be a leaf  species such that the edge that connects $s$ and its parent node $s'$ in $T$ is  labeled by the inactive character ${d}^-$ (i.e. $d$ is not in the root of tree $T$).
	Let $c$ be the unique  character in   species $s$  which is not in the  initial species of the tree $T$. Since the tree $T$ is \incomplete (i.e. is not \simple) by Lemma \ref{lem:single-source-more-characters} the initial species of tree $T$ has the only character $b$.
	Observe that species  $s'$   follows $s$ in the incomplete chain $\cal C$ induced by $s, s'$ in the poset. Now $s'$ contains the character $d$ that is negated on the edge from $s'$ to $s$ in the tree $T$.
	The following cases must be considered: case 1) $d$ is gained along the initial path before $c$, case 2)  $c$ and $d$ are gained together, case 3) $c$ is gained before  $d$.
	Case 1). Let $e$ be another character gained along the initial path (indeed, the tree must have at least four characters) on an edge distinct from the one of $d$.  Now since $s$ is a leaf with the single character $c$, then  $d$ and $e$ must be negated along the same branch of the umbrella in the same order they are introduced,  by the main property Proposition~\ref{prop:basic-1}. 
	If $s'$ occurs in the branch before the leaf, then $d$ is negated after $e$ is negated, i.e. $e$ is gained before $d$, and thus since $c$ is gained after $d$, it follows that $e$ is gained before $c$ and before $d$ along the initial path.  But this fact will lead to contradict the  Lemma~\ref{lem:property-inverted-order}, by which there exists a pair of characters such that there are lost along distinct branches of the umbrella. Indeed by the generality of characters $e$ any character before $d$ is negated along the same branch of $d$.  Observe that it must be that $b$ is negated along the initial path of tree $T$ since $T$ is \incomplete.  
	Thus, consider case 2), that is  $c, d$ are gained together.  
    Then $e$ is negated in the tree below the edge gaining  $c, d$ together. Now, there exists a species $s''$ that follows $s'$ in the chain having also $e$. But, we can show that $(c, e)$ is a conflicting pair in such species $s''$. Indeed,   there exists two species with $c$ and without $e$: one is the leaf $s$ and the other is the species $s'$ after the loss of $e$ in the umbrella. 
    Similarly, there is a species with $e$ and not $c$ which is the one in the umbrella below the negation of $c$.
    Indeed, $c$ and $e$ must be negated along distinct branches of the tree  in virtue of Proposition~\ref{prop:basic-1}. 	
    It follows that case 2) is not possible since the chain $\cal C$ cannot be active having species $s''$ whose realization induces a red $\Sigma$-graph having a conflicting-pair by Lemma \ref{lem:conflict-pair}.
	Case 3). Assume that $c$ is gained before $d$. Then we can apply Lemma~\ref{lem:technical-01} which shows that the chain is not active as it starts with the pair $(s,s')$. 
 Also, this case is not possible, thus concluding the proof of the Lemma.
\end{proof}

Let us first state a technical Lemma used to prove Lemma~\ref{lem:incomplete-chain-with-leaf}.
\begin{lemma}
	\label{lem:technical-01}
	Let $\grbm$ be a skeleton graph and let $T$ be a tree solving the graph. Let $s$ be a leaf  species such that the edge that connects $s$ and its parent $s'$ in $T$ is  labeled by the inactive character ${d}^-$.
	Let $c$ be an inactive character in  $C(s)$  which is not in the  initial species of the tree $T$ and $c$ is gained before $d$ along the initial path of tree $T$. Then the realization of $(s, s')$  induces the red-$\Sigma$ graph.
\end{lemma}

\begin{proof}
	We can show that the pair  $(c, d)$ is a conflicting pair  in the species $s'$ of graph $\grbm$ obtained after the realization of species $s$,  thus implying that the realization of the pair $(s, s')$ leads to a red-$\Sigma$ graph.
	By hypothesis, $c$ and $d$ occur below the initial species $s_0$ and then $s_0$ induces the configuration $(0,0)$  for $c$ and $d$  (indeed $c$ is not in $C(s_0)$). Moreover,  species $s'$ has the configuration $(1,1)$ for both pair of characters. Let $s_1$ be the species where $c$ is gained in the tree. Then there are at least two  species, precisely, $s_1$  and $s$ having $c$ and not $d$, which implies that the pair $(c, d)$ induces the $(1,0)$ configuration in those two species and $s_1$ is still present in the graph $\grbm$ after the realization of $s$.  Moreover the configuration $(0,1)$ is induced by the pair $(c, d)$ in a species of the umbrella of tree $T$,  that is in a branch where $c$ must be negated (or lost): such a species exists since $c$ and $d$ are not comparable characters and $c$ occurs before $d$ by hypothesis (see \Cref{prop:basic-1}). Thus after the realization of $s$,  $(c, d)$ are characters both present in $s'$ thus showing that the realization of $s'$ leads to a red $\Sigma$-graph thus completing the proof.
\end{proof}

{\bf Proof of \Cref{lem:property-branch-incomplete}}
{\em Let $\grbe$ be an incomplete skeleton graph solved by a branch-tree.
	Then the graph has only one active   chain and its source $s$ is the only safe species in the graph.}

\begin{proof}
We first prove the Lemma for the skeleton graph \grbm  of \grbe and then show that the Lemma extends to \grbe.
Indeed, given $T_e$	the tree that solves the graph \grbe and $T$ the  branch-tree solving the graph $\grbm$, it holds that since the graph  \grbm has only one active chain whose source is the only  safe species, the same property extends to \grbe. Indeed, observe that any  active chain in \grbe is also active in \grbm. But graph \grbe will have at least an active chain since it is solved by a tree and thus the unique active chain in \grbm is the active one in \grbe.   
	Since $T$ is \incomplete,  by definition \ref{def:complete grb} it means that tree $T$  does not have a species that has all inactive characters. Consequently, the initial path  of tree $T$ contains the negation of an
	inactive character  which is the first to be introduced in virtue of Proposition~\ref{prop:basic-1}. Consider the chain ${\cal C}$ obtained by enumerating the species of the tree $T$ starting from the source of $T$ to the species before the  edge where the first negation of an inactive character occurs.
	Clearly, ${\cal C}$ is an incomplete chain   whose realization in graph \grbm leads to a reducible graph (in vitue of Proposition \ref{prop:traversal}), whose source is a safe species in \grbm, i.e. ${\cal C}$ is an active incomplete chain  in graph \grbm.
	Now, let us show that $\cal C$ is the unique active \incomplete  chain, thus proving what required.
	By Lemma~\ref{lem:nosource-is-internal-species} any source of an active chain is either the initial state or the leaf of a tree $T$ solving the skeleton graph thus implying that the only other possible active chain starts with a leaf of the branch-tree $T$.
	Let us now show that this is not possible.
	Assume to the contrary that $s$ is the leaf of the branch-tree $T$ and the source of an active chain $\cal C'$.  

	Observe that by Lemma~\ref{lem:single-source-more-characters} the initial state must contain a unique character, assume $b$, that is the first to be negated by Proposition~\ref{prop:basic-1}. 
    In the following we apply this observation: as a consequence of the fact that the initial state induces the $(0,0)$ configuration for any pair of characters $(c,c')$ occurring below the initial state, the pair $(c,c')$  is conflicting in a species having both characters. Indeed $c,c'$  are not comparable being maximal characters and the realization of the pair of characters in any order does not remove species, by definition of feasible species.	
    Since the  inactive character $b$ is negated along the initial path, it means that either:  ii.1) leaf $s'$ of the branch-tree  contains  at least  two inactive characters $c, d$ of  the initial path of tree $T$ that are not in the initial state of tree $T$,    ii.2)  leaf $s'$ consists of a single inactive character that is not in the initial state of tree $T$.
	Assume that case  ii.1)  holds. Observe that when realizing $s'$ by definition it must be a minimal species. Then by the initial observation, $c, d$  is a conflicting-pair in   $s'$ when $s'$ is realized which means that the realization of $c$ and $d$ induces the red $\Sigma$-graph in the graph.  It follows  that  $\cal C'$ is not an active chain,  a contradiction. Thus case ii.1) is not possible.
	Assume that case  ii.2)  holds.  Since $s$ is a leaf  species  having  a single inactive character $c$ that
	is not in the  initial species of the tree $T$, we can apply Lemma~\ref{lem:incomplete-chain-with-leaf} showing that $s'$ cannot be the source of an active chain, and thus  $\cal C'$ is not an active chain, a contradiction. Both cases ii.1) and ii.2) lead to a contradiction, showing that case ii) is not possible. This fact  completes the proof of the Lemma.
\end{proof}


{\bf Proof of Lemma~\ref{prop:property-line-sufficient}}
{\em Let $\grbm$ be a skeleton  graph that is solved by a line-tree.
	Then the graph has only two active chains that may be both incomplete  or complete.
	The two active chains have sources that correspond to the initial state of two trees $T$ and $T'$ solving the graph where $T'$ is the inverted-tree of $T$.}
\begin{proof}
	Given a line-tree solving the graph, we may have two situations, either all inactive characters are introduced  and then negated one after the other and then the graph has at most two active complete chains (with $A$ empty), or active chains in the graph are incomplete  which means that characters are negated along the initial path making the chains incomplete.
	By lemma	\ref{lem:nosource-is-internal-species} any source of an active chain is either the initial state or a leaf of the tree.
	In a line-tree there is only one initial state and one leaf leading to prove that there are only two complete or  incomplete active chains and it is immediate that their sources are safe since the leaf is the initial state of an inverted-tree of the original tree $T$.
	Thus, there are at most two trees solving the graph each one starting with one of the two active chains, whose sources are the initial state of these two trees.
\end{proof}


\section{Conclusion}
Relaxing the infinite sites assumption in the Perfect Phylogeny model, while retaining its computational efficiency, remains a major open problem in tumor phylogeny reconstruction. A natural first step is to allow persistent characters, as formalized by the Persistent Phylogeny (PP) problem, which generalizes the Dollo-1 model by permitting each character to be lost at most once in the tree.

In this work, we provide a polynomial-time solution to the PP problem, thereby resolving a longstanding question related to constructing a tree that explains a binary matrix of species and characters. Notably, the PP problem is equivalent to the Generalised Character Compatibility (GCC) problem in the non-branching case, an open problem posed nearly 30 years ago in \cite{benham95}. Restricted variants of the GCC problem have also been studied in \cite{mavnuch2009generalised}, where the general Dollo-1 case corresponds to the open problem (5) illustrated in Fig. 1.

Our algorithm is based on a graph representation of PP problem instances and a detailed characterization of trees that solve special cases represented by skeleton graphs. In particular, we show that for each such graph, there are at most two distinct solutions unless the graph is degenerate. This characterization reveals that solutions for skeleton graphs have a special structure, which also constrains the possible tree constructions for general instances.

The polynomial-time algorithm we present hinges on relating the properties of trees that solve the red-black graphs to the process of reducing these graphs. However, an open question remains: whether there exists a polynomial-time algorithm that can recognize reducible red-black graphs directly from their structural properties, similar to the approach taken for the Incomplete Perfect Phylogeny problem~\cite{Sha}.

It is worth noting that the red-black graph introduced in \cite{DBLP:journals-tcs-BonizzoniBDT12} provides a graph formulation of the PP problem on specific incomplete matrices. Moreover, a complete characterization of forbidden substructures for reducible red-black graphs is still unknown. The results discussed in this paper could provide valuable insights into these two open questions.

\section{Acknowledgment}
The authors would like to thank Murray Patterson for helpful discussion on the topic.
P.B and G.DV. have received funding from the European Union’s Horizon 2020 Innovative Training Networks programme under the Marie Skłodowska-Curie grant agreement No. 956229.

P.B and G.DV. have received funding from the European Union’s Horizon 2020 Research and Innovation Staff Exchange programme under the Marie Skłodowska-Curie grant agreement No. 872539.

\bibliographystyle{abbrv}
\bibliography{books,persistent,tumor,references}

\end{document}